\documentclass[twocolumn,noshowpacs,nofootinbib,superscriptaddress]{revtex4}

\pdfoutput=1

\usepackage{enumitem}
\usepackage{amsmath,amssymb,bm,dsfont}%,mathtools}
\usepackage{graphicx}

\newcommand{\grad}[2]{\partial_{#2} #1 }

\newcommand{\vu}{\bm{u}}
\newcommand{\vf}{\bm{f}}
\newcommand{\vv}{\bm{v}}

\newcommand{\vm} {\bm{m}}

\newcommand{\vR}{\bm{R}}

\newcommand{\unitl}{{l_0}}
\newcommand{\unitk}{{k_0}}
\newcommand{\uniten}{\unitk{\unitl}^3}
\newcommand{\unitforce}{\unitk{\unitl}^2}
\newcommand{\unitfreq}{\omega_0}
\newcommand{\unitc}{{c_0}}
\newcommand{\unitmom}{{m_0}}
\newcommand{\unitmag}{{M_0}}
\newcommand{\celsius}{^\circ\mbox{C}}

\newcommand{\np}{{4631}}
\newcommand{\ndof}{{13893}}
\newcommand{\nmagn}{{771}}
\newcommand{\nmesh}{{3860}}

\newcommand{\threshold}{{10^{-3}}}
\newcommand{\Nsamples}{15}
\newcommand{\Nsprings}{50400}

\newcommand{\e}{\textrm{e}}
\newcommand{\imu}{\textrm{i}}

\newcommand{\sumtwolines}[3]{\sum_{\substack{#1{}\\#2}}^{#3}}

\usepackage{ulem, color}
\newcommand{\GP}[1]{\textcolor{black}{#1}}

\begin{document}

%\onecolumngrid
%\begin{flushleft}
%{\it Journal Reference: \\
%URL: \\
%DOI: \\}
%\end{flushleft}
%\twocolumngrid

\title{Tunable Dynamic Moduli of Magnetic Elastomers: \\ From X-$\mu$CT Characterization to Mesoscopic Modeling}

\author{Giorgio Pessot}
\email{giorgpess@uni-duesseldorf.de}
\affiliation{Institut f\"ur Theoretische Physik II: Weiche Materie,
 Heinrich-Heine-Universit\"at D\"usseldorf, D-40225 D\"usseldorf, Germany}

\author{Malte Sch\"umann}
\email{malte.schuemann@tu-dresden.de}
\affiliation{Technische Universit\"at Dresden, Institute of Fluid Mechanics, D-01062, Dresden, Germany}

\author{Thomas Gundermann}
\email{thomas.gundermann@tu-dresden.de}
\affiliation{Technische Universit\"at Dresden, Institute of Fluid Mechanics, D-01062, Dresden, Germany}

\author{Stefan Odenbach}
\email{stefan.odenbach@tu-dresden.de}
\affiliation{Technische Universit\"at Dresden, Institute of Fluid Mechanics, D-01062, Dresden, Germany}

\author{Hartmut L\"owen}
\email{hlowen@uni-duesseldorf.de}
\affiliation{Institut f\"ur Theoretische Physik II: Weiche Materie,
 Heinrich-Heine-Universit\"at D\"usseldorf, D-40225 D\"usseldorf, Germany}

\author{Andreas M.\ Menzel}
\email{menzel@.thphy.uni-duesseldorf.de}
\affiliation{Institut f\"ur Theoretische Physik II: Weiche Materie,
 Heinrich-Heine-Universit\"at D\"usseldorf, D-40225 D\"usseldorf, Germany}

\date{\today}

\begin{abstract}
Ferrogels and magnetoelastomers are composite materials obtained by embedding magnetic particles of mesoscopic size in a crosslinked polymeric matrix.
They combine the reversible elastic deformability of polymeric materials with the high responsivity of ferrofluids to external magnetic fields.
\GP{These materials} stand out, for example, for \GP{large} magnetostriction as well as \GP{significant} increase of the elastic moduli in the presence of external magnetic fields.
By means of X-ray micro-computed tomography, position and size of each magnetic particle can be measured with a high degree of accuracy.
We here use data extracted from real magnetoelastic samples as input for coarse-grained dipole--spring modeling and calculations to investigate magnetostriction, stiffening, and changes in the normal modes spectrum.
More precisely, we assign to each particle a dipole moment proportional to its volume and set a randomized network of springs between them that mimics the behavior of the polymeric elastic matrix.
\GP{Extending our previously developed methods, we compute the resulting structural changes in the systems, their overall distortions, as well as the frequency-dependent elastic moduli when magnetic interactions are turned on.
Particularly, with increasing magnetization, we observe the formation of chain-like aggregates, resulting in significant overall deformations.
Interestingly, the static elastic moduli can first show a slight decrease with growing amplitude of the magnetic interactions, before a pronounced increase appears upon the chain formation.
The change of the dynamic moduli with increasing magnetization depends on the frequency  and can even show nonmonotonic behavior.}
Overall, we demonstrate how theory and experiments can complement \GP{each other} to learn more about the dynamic behavior of this interesting class of materials.
\end{abstract}

\pacs{82.35.Np, 63.50.-x, 62.20.de, 75.80.+q}

% 82.35.Np    Nanoparticles in polymers (see also 81.07.-b Nanoscale materials and structures: fabrication and characterization)
% 63.50.-x    Vibrational states in disordered systems
% 62.20.de    Elastic moduli
% 75.80.+q    Magnetomechanical effects, magnetostriction (for magnetostrictive devices, see 85.70.Ec)

% 83.60.Bc    Linear viscoelasticity
% 62.20.D-    Elasticity
% 62.20.Hg    Creep
% 82.70.Dd    Colloids
% 02.60.Dc    Numerical linear algebra
% 02.60.Pn    Numerical optimization
% 82.70.Gg    Gels and sols
% 82.70.-y    Disperse systems; complex fluids (see also 82.33.-z reactions in various media; for quantum optical phenomena in dispersive media, see 42.50.Nn)

\maketitle

\section{Introduction}\label{introduction_section}
One powerful method to devise innovative, enhanced materials is to combine two or more known components into a new substance featuring new and optimized properties.
Among composite materials, magnetoelastic materials such as ferrogels and magnetic elastomers \cite{filipcsei2007magnetic, menzel2015tuned, lopez2016mechanics, odenbach2016microstructure, asun2017modeling} blend the elastic, reversible deformability typical of polymeric materials \cite{strobl1997physics, doi1988theory} with the responsiveness to external magnetic fields distinctive of ferrofluids \cite{odenbach2003ferrofluids, huke2004magnetic}.
Their possible applications include soft actuators \cite{zimmermann2006modelling, schmauch2017chained}, tunable vibration absorbers \cite{deng2006development, sun2008study}, tunable damping devices \cite{abramchuk2007novel}, magnetic field detectors \cite{szabo1998shape,ramanujan2006mechanical}, electromagnetic radiation absorbers \cite{sedlacik2016magnetorheological}, and smart acceleration sensors \cite{volkova2017motion, becker2017dynamic}.
Since the polymer can be bio-compatible, applications for drug delivery \cite{zhou2015magnetic} have been outlined.

\GP{One way to generate soft magnetoelastic materials is to crosslink a polymeric solution in which magnetic particles of mesoscopic size are dispersed.}
The particles are typically large enough \GP{so that they cannot move through the surrounding} polymer mesh, or they are directly chemically bound to the polymer network \cite{frickel2011magneto, bose2015relationship, landers2015particle}.
Therefore, when magnetically interacting with an external field and with each other, the particles \GP{rotate or push} against their environment and in this way deform the surrounding polymer matrix.
Thus the resulting ``magneto-mechanical'' coupling \cite{frickel2011magneto, kaestner2013higher, allahyarov2014magnetomechanical} can reversibly change the shape and stiffness of the materials in the presence of an external magnetic field.
Consequently, \GP{adjusting the magnetic interactions by external magnetic fields, it is possible to tune} the mechanical state and properties such as strain \cite{zrinyi1996deformation, zrinyi1997ferrogel, guan2008magnetostrictive, gong2012full, varga2005electric} or elastic moduli \cite{jolly1996magnetoviscoelastic, varga2006magnetic, stepanov2007effect, kallio2007dynamic, bose2009magnetorheological, evans2012highly, borin2013tuning, schubert2016equi, schuemann2017insitu}.
This coupling can also influence other physical properties such as resonance frequency \cite{volkova2017motion}, or electrical impedance \cite{wang2016stimuli}.
Furthermore, it is the origin of intriguing features such as formation of chains of magnetic particles \cite{schuemann2017insitu} and their buckling under magnetic fields \cite{huang2015buckling}, superelasticity \cite{cremer2015tailoring, cremer2016superelastic}, and complex behaviors of the  dynamic moduli \cite{pessot2016dynamic}.

Various theoretical approaches have been performed to describe the behavior of these materials, such as macroscopic theory \cite{szabo1998shape, jarkova2003hydrodynamics, bohlius2004macroscopic, allahyarov2014magnetomechanical, menzel2014bridging}, finite-element methods \cite{stolbov2011modelling, han2013field, metsch2016numerical, romeis2017theoretical, kalina2017modeling}, as well as particle resolved models \cite{ivaneyko2011magneto, weeber2012deformation, pessot2014structural, cremer2015tailoring, cremer2016superelastic, pessot2016dynamic}.
One new route in the modeling has recently been outlined in the form of a density functional theory \cite{cremer2017density}.
Along these lines, the magnetically induced  changes in elastic properties have been described \cite{ivaneyko2011magneto, ivaneyko2012effects, han2013field, pessot2014structural, cremer2015tailoring, cremer2016superelastic, pessot2016dynamic}.

In the particle-resolved approaches, the magnetic particles are usually assumed to carry magnetic point dipoles, a reasonable approximation at low volume fractions \cite{biller2014modeling}.
The challenge typically consists in representing appropriately the elastic forces mediated by the polymer matrix.
To lowest order, the matrix-mediated elastic interaction can be described via linear \cite{annunziata2013hardening, cerda2013phase, pessot2014structural, pessot2016dynamic} or nonlinear \cite{sanchez2013effects} springs connecting the particles.
Other approaches coarse-grain the polymer into a coupled \GP{mesh of} nearly-incompressible tetrahedra \cite{cremer2015tailoring, cremer2016superelastic}, or, conversely, zoom onto the microscopic detail by resolving the individual polymer chains in a coarse-grained way \cite{pessot2015towards, weeber2015ferrogels}.
Within the framework of linear elasticity theory, the matrix-mediated interactions between the particles can be calculated analytically and up to a desired order exactly \cite{puljiz2016forces, puljiz2017forces, menzel2017force, kim1994faxen, phan-thien1994load}.

When devising new materials with smart, enhanced properties, one usually aims at optimizing their response to an external perturbation.
For instance, in the case of a vibration absorber, one would like to have a material stiff enough to sustain the required load but also viscous enough to dissipate kinetic energy as \GP{quickly} as possible.
The key physical \GP{quantities are} the dynamic moduli.
They provide the time- or frequency-dependent stress response of the material to an externally imposed strain.
Recently, increasing attention has been paid on investigating the time-dependent properties either via macroscopic \cite{nadzharyan2016fractional, jarkova2003hydrodynamics, bohlius2004macroscopic} or particle-resolving \cite{tarama2014tunable, ivaneyko2015dynamic, pessot2016dynamic} models.

In our previous investigation \cite{pessot2016dynamic} we outlined a method to compute the frequency-dependent Young and shear moduli $E(\omega)$ and $G(\omega)$, respectively, for basically any given particle arrangement and studied several theoretically assumed particle structures.
In the present work, after refining our method, we aim at applying our \GP{technique to three-dimensional} ($3$D) particle distributions obtained by X-ray micro-computed tomography (X-$\mu$CT) \cite{schuemann2017insitu, gunther2012xray, borbath2012xmuct, gundermann2013comparison} from real experimental samples.
We evaluate the dynamic moduli for varying frequencies, volume fractions of magnetic particles, and intensity of the magnetic interactions.

Our paper is structured as follows.
First, in section \ref{model_section} we present our minimal dipole--spring model including steric repulsion.
Then, in section \ref{micro_tomo_section}, we describe the experimental set-up employed to acquire the particle distributions used as input for our model.
Then, our technique to calculate the dynamic moduli is briefly summarized in section \ref{dyn_mod_calc_section}.
Finally, we present our results in section \ref{results_section} before drawing our final conclusions in section \ref{conclusions_section}.

\section{Dipole--Spring Model}\label{model_section}
Our employed dipole--spring model is \GP{a modified version of} the one presented in Ref.~\onlinecite{pessot2016dynamic}.
First, we summarize the properties of our particle--spring network and explain how it is generated.
Then, we \GP{discuss} the pair potentials acting between the particles and the generated mesh nodes in the reduced units of the system.

\subsection{Particle--Spring Network}\label{model_intro_section}
We consider a set of $N_p$ spherical magnetic particles with positions $\vR_i = (x_i, y_i, z_i)$ and radii $a_i$ ($i=1\dots N_p$).
The positions and radii of the magnetic particles are experimentally measured and used as input for our calculation (see section \ref{micro_tomo_section} below).
Moreover, each particle carries an individual magnetic moment $\vm_i$.
We assume an identical magnetization $\bm{M}$ for all particles as might be achieved by magnetization \GP{in a homogeneous} external magnetic field, both under magnetic saturation or when neglecting mutual particle magnetization.
Together with the particle volume $v_{p,i}=\frac{4}{3}\pi a_i^3$ we thus obtain $\vm_i =\bm{M} v_{p,i}$.

The polymeric network embedding the particles is modeled by a network of harmonic springs.
In our previous works we introduced a minimal set of springs directly connecting the particles via \GP{Delaunay} triangulation \cite{borbath2012xmuct, gunther2012xray, tarama2014tunable, pessot2014structural}.
Care was taken to suppress artificial soft shear modes.
However, in the present work, to represent \GP{in a better way} a uniform elastic background around irregular particle arrangements we insert $N_{extra}$ extra nodes in the space between our magnetic particles.
These nodes are non-magnetic, volumeless, and only serve as linking points connecting springs.
We similarly label the positions of these mesh nodes by $\vR_i = (x_i, y_i, z_i)$, with $i =N_p+1, \dots, N$. 
The $N_p$ magnetic particles and the $N_{extra}$ additional nodes add up to a total of $N=N_p +N_{extra}$ mesh points, implying $3N$ translational degrees of freedom.

The nodes of the total network are initially positioned according to a face-centered-cubic (fcc) lattice.
Each lattice site corresponds to a mesh node.
The nearest-neighbor connecting edges between the nodes are converted to harmonic springs.
Furthermore, $a_{mesh}$ sets the initially identical length of all springs.
We choose an appropriately small mesh size comparable to the average interparticle distance in the densest sample, see section \ref{micro_tomo_section}.
In other words, we set our spring network so that it ``fills'' the gaps between the magnetic particles as homogeneously as possible.
Moreover, $a_{mesh}$ is identical for the different investigated systems.

\begin{figure}
\centering
  \includegraphics[width=8.6cm]{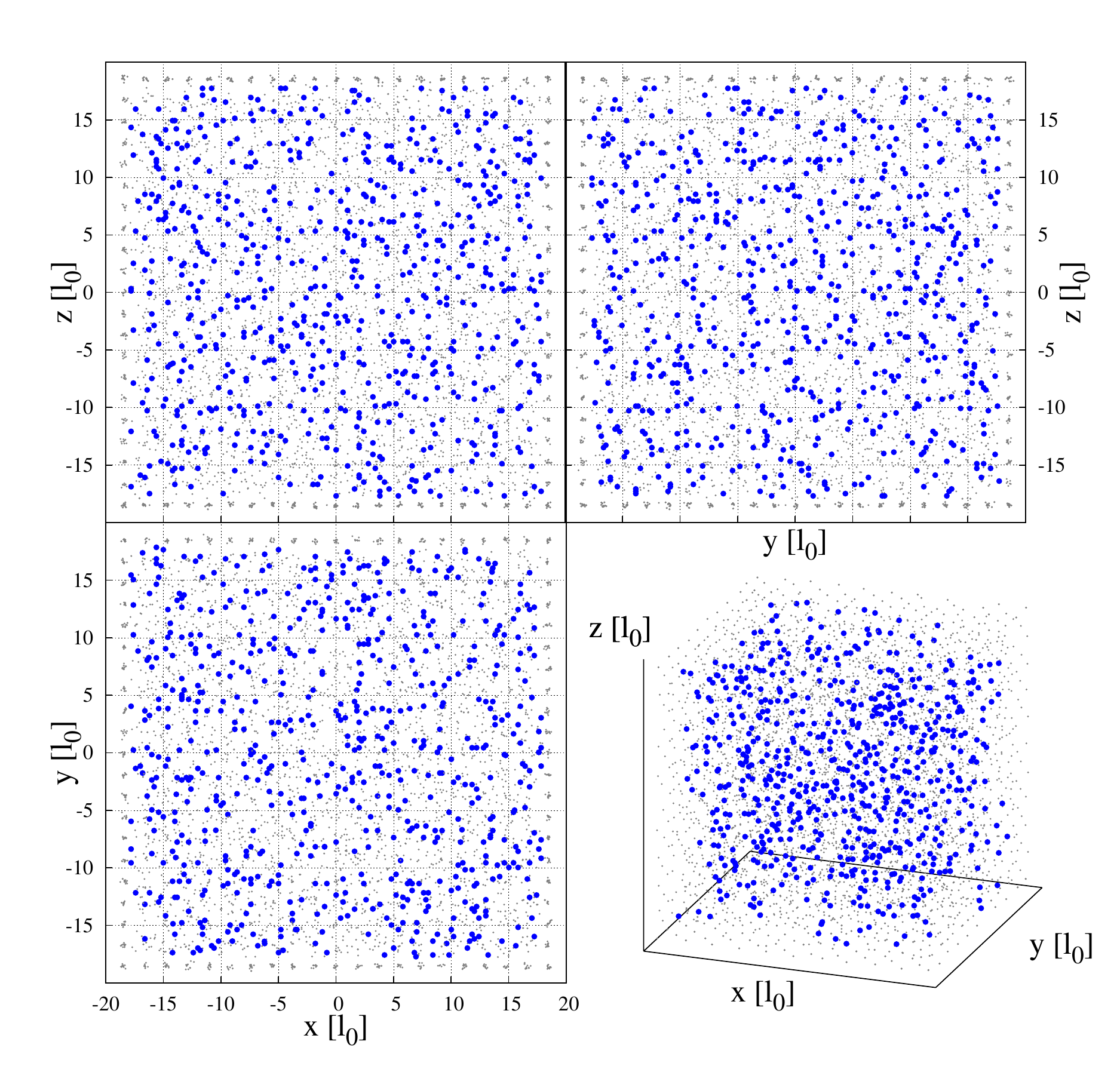}
  \caption{Cartesian projections and $3$D plot of magnetic particles (larger, blue spheres) and mesh nodes (smaller, gray spheres) in one numerical realization of an experimental sample.
  The mesh nodes are numerically generated to represent a homogeneous polymeric matrix.
  Experimentally, the arrangement of magnetic particles is obtained via X-$\mu$CT (see section~\ref{micro_tomo_section}).
  A central, homogeneous region of the sample is then cut for the numerical system generation (here of the $40$ wt\% sample), see section \ref{micro_tomo_section}.}
  \label{part_mesh_15}
\end{figure}

When later an experimentally measured particle configuration of overall cubic shape is imported, it is laid over the spring network.
Subsequently, the network node closest to each particle is moved onto its center and the particle is attached to it.
We determine the mean square displacement of all these \GP{displacements}.
In a subsequent step, the remaining nodes are stochastically displaced with an identical mean squared displacement, thus randomizing the whole network homogeneously, see Fig.~\ref{part_mesh_15}.
The lengths of the springs in their undeformed states are adjusted accordingly, so that there are no initial stresses in the elastic network in the absence of magnetic interactions. To maintain an overall cubic shape with well-defined boundaries, the nodes on the outermost layers of the network are, however, \GP{only} randomized by \GP{one fifth of that} amount.

It is important to define the boundaries of the resulting configuration, i.e., the faces of the cubic-like system.
There are six boundaries: front, rear, left, right, top, and bottom, corresponding, respectively, to the $\pm\widehat{\bm{y}}$, $\pm\widehat{\bm{x}}$, and $\pm\widehat{\bm{z}}$ surfaces.
In the following, we will denote the set of particles and nodes belonging to each boundary by $\mathcal{B}_{\widehat{\bm{\alpha}}}$ ($\widehat{\bm{\alpha}}=\pm\widehat{\bm{x}}, \pm\widehat{\bm{y}}, \pm\widehat{\bm{z}}$).
The boundaries of our system are chosen in a way to include about the same number of particles as there are on the faces of the initial fcc network.
Later, the magnetized particles will be displaced under increasing magnetization.
The particles assigned to the boundaries can change but their  overall number remains constant.
Below, we will be employing a total of $N = \np$ nodes.
Each boundary comprises about $\sim 5\%$ of them.

Each spring connecting the $i$-th mesh point (particle or extra node) with the $j$-th one is characterized by its elastic constant $k_{ij}$.
We set the elastic constant of each spring to be proportional to its length in the undeformed state.
In a simplified picture, the overall elastic constant for stretching a cylinder or cuboid of elastic material along its axis scales like the ratio of its cross section over its length.
In the case of diluted dispersions of particles in an elastic medium, the cross-section of the material in between them will scale like the square of their distance.
Therefore, and as a first approximation, the elastic constant of their elastic interaction will be proportional to their distance.
Furthermore, the resulting overall elastic modulus of the whole system should not depend on the artificially introduced length scale $a_{mesh}$.
Indeed, we verified that scaling each spring constant proportional to its length yields converging moduli for decreasing $a_{mesh}$.

Finally, we assume an overdamped dynamics of our particles and extra mesh nodes \cite{tarama2014tunable, ivaneyko2015dynamic, pessot2016dynamic}.
Therefore, the motion of each particle and extra node is subject to an effective frictional force $-c_i \dot{\vR}_i$, with the coefficients $c_i = \unitc$ identical for each node, both for particles and extra mesh node.
The reason is that here the relaxation process is mainly determined by the bulk elastic medium, which, in both cases, is given by the same (visco)elastic \GP{polymer.}
\GP{An additional solvent can further modify the dynamic relaxation behavior.}

\subsection{Pair Interactions}\label{pair_potentials_section}
We now detail the various contributions to the total energy of the system $U$.
It is composed of elastic $U^{el}$, magnetic $U^{m}$, and steric $U^{s}$ contributions such that $U=U^{el}+U^{m}+U^{s}$. 
First, the elastic energy $U^{el}$ stored in the elastic springs reads
\begin{equation}\label{eel}
 U^{el}= \frac{1}{2}\sumtwolines{i,j=1}{i \neq j}{N} \frac{k_{ij}}{2} {\left( r_{ij} -\ell_{ij}^0 \right)}^2,
\end{equation}
where the sum runs over all mesh nodes (particles and extra nodes) labeled by $i$ and $j$ ($j\neq i$).
Furthermore, $\bm{r}_{ij}=\vR_j -\vR_i$, $r_{ij}=|\bm{r}_{ij}|$, and $\ell_{ij}^0$ is the length of the spring connecting mesh nodes $i$ and $j$ in the undeformed state. 
The spring constants are given by $k_{ij} = \unitk \ell_{ij}^0$ (see above), if $i$ and $j$ are connected by a spring, and $0$ otherwise.
$\unitk$ then sets the moduli of the overall matrix.

\GP{Since we assume the magnetization $\bm{M}$ constant for all particles along an identical direction, set, e.g., by an external magnetic field, the only contribution to the magnetic energy $U^m$ due to varying particle distance is the dipole--dipole interaction}
\begin{equation}\label{emagn}
 U^{m}= \frac{\mu_0}{4\pi} \ \frac12 \sumtwolines{i,j=1}{i \neq j}{N_p} \frac{\vm_i\cdot\vm_j r_{ij}^2 -3{(\vm_i\cdot\bm{r}_{ij})}{(\vm_j\cdot\bm{r}_{ij})}}{r_{ij}^5}.
\end{equation}
Here, the sum runs only over the pairs of the $N_p$ magnetic particles, $\vm_i=\bm{M} v_{p,i}=\bm{M} \frac43 \pi {a_i}^3$, and $\mu_0$ is the magnetic permeability of vacuum.

Finally, the steric repulsion is given by
\begin{equation}\label{est}
 U^{s}=\frac{1}{2} \sumtwolines{i,j=1}{i \neq j}{N_p} v^s\!\left(\frac{r_{ij}}{\sigma_{ij}^s}\right),
\end{equation}
where $v^{s}$ is a purely repulsive interaction.
It is given by
\begin{align}
v^{s} (\xi)&=\varepsilon^{s} \Bigl[ {\xi}^{-12}- {\xi}^{-6} +{\varepsilon_c} - \frac{c^{s}}{2} {(\xi -\xi_c)}^2 \Bigr]
\end{align}
for $\xi<\xi_c$ and zero otherwise \cite{pessot2016dynamic}, in the spirit of the WCA \cite{weeks1971role} potential.
The shift $\varepsilon_c$ and the cutoff range $\xi_c =2^{1/6}$ are set to achieve $v^{s} \left(\xi_c\right) = 0$ and $ v^{s\prime} \left(\xi_c\right) =0 $. 
Moreover, \GP{the term with the coefficient $c^s$ is introduced to achieve $v^{s \prime \prime} \left(\xi_c\right) = 0$, which requires $c^s = 36\times 2^{-4/3}$ (see appendix A of Ref.~\onlinecite{pessot2016dynamic} for further details).}
Since we here consider particles of varying sizes, we arrange the steric cutoff between particles $i$ and $j$, where steric interaction sets in, to correspond to the sum of the respective radii.
Thus, $\sigma^s_{ij}=2^{-1/6}(a_i+a_j)$.
%Additionally, we choose $\varepsilon^s=1$ in our reduced units.

To further describe our systems below, we define the characteristic length scale $\unitl=\sqrt[3]{\sum_{i=1}^{N_p} a_i^3/N_p}$.
Thus, the total volume occupied by the particles is exactly $N_p\frac{4}{3}\pi\unitl^3$.
Moreover, and as further described in section \ref{micro_tomo_section}, the distributions of particle radii in the considered samples have similar averages.
Thus $\unitl$ is a good length scale to compare systems of different particle concentration. We set the strength of our steric interaction to $\varepsilon^s=\unitk\unitl^3$.

\section{Experimental Data Acquisition and Characteristics of the Resulting Numerical Systems}\label{micro_tomo_section}
The experimental samples from which we acquired the particle positions are of cylindrical shape and with weight percentage (magnetic particles over elastomer weight ratio) of $15$ and $40$ wt\%.
Manufacturing processes, data acquisition, \GP{as well as a comprehensive evaluation of measured particle structures and mechanical properties of the samples are given} in Refs.~\onlinecite{gundermann2014investigation, gundermann2017statistical} ($15$ wt\%), and \onlinecite{schuemann2017insitu} ($40$ wt\%).
In size, the samples had diameters of $3.5\mbox{ mm}$ and $4\mbox{ mm}$ as well as heights of $3.5\mbox{ mm}$ and $5\mbox{ mm}$, respectively.
To prepare the polymer host matrix, the elastomer kit Elastosil$^{\circledR}$ RT 745 A/B (Wacker Chemie AG, Germany) was employed for the sample with $15$ wt\% and silicone polymers by Gelest Inc.\ for the sample of $40$ wt\%.
%Their composition consists of Polydimethylsiloxane DMS-V25 (PDMS) mixed with Copolymer HMS-151 and catalyst in proportion of $2.8$ and $0.12$ parts every $100$ parts of PDMS, respectively.
%The catalyst employed is Alfa Aesar Platinum (0)-1,3-divinyl-1,1,3-tetramethyldisiloxane, 1:20.

In both cases, soft-magnetic carbonyl iron powder ASC200 (H\"ogan\"as AB, Sweden) was added.
%The average size of the magnetic particles was about $35-40\ \mu\mbox{m}$, see also Ref.~\onlinecite{gundermann2014investigation, schuemann2017insitu}.
The amounts of iron powder of $15$ and $40$ wt\% (volume fraction $\phi \simeq 0.023$ and $\simeq 0.056$) were chosen to obtain significant responsiveness to magnetic fields as well as a statistically significant amount of particles described.
On the one hand, the $15$ wt\% sample was prepared by pouring the silicon-iron powder mixture into a mold which was then immersed into a water bath at $95\ \celsius$ for two hours for polymerization.
On the other hand, the $40$ wt\% sample was polymerized by action of the catalyst (Alfa Aesar Platinum (0)-1,3-divinyl-1,1,3-tetramethyldisiloxane, 1:20) with a short final phase ($30$ minutes at $100\ \celsius$) of high-temperature curing to finish.
\GP{Care was taken to avoid particle sedimentation and to ensure a homogeneous distribution of the particles in the polymerized sample.}

In a successive stage, X-ray micro-computed tomography (X-$\mu$CT) scans of the samples were performed.
\GP{An X-$\mu$CT system \cite{shevchenko2013application}} with electron current and acceleration voltage \GP{set to} $170\ \mu\mbox{A}$ and $90\ \mbox{kV}$, respectively, was employed.
Projected images of the samples were generated by rotating the sample with a $0.25^\circ$ increment.
Furthermore, throughout the CT investigations, temperature remained constant \GP{at $20\ \celsius$.}
The exposure time, instead, varied from $2\ \mbox{s}$ to $6.5\ \mbox{s}$ to optimize the quality of the resulting image, for which a magnification factor of $15$ was used, thus achieving a resolution of $1\ \mbox{pixel}=3.2\ \mu\mbox{m}$.
Finally, a self-developed software based on the FDK algorithm \cite{feldkamp1984practical} was used \GP{to reconstruct the $3$D images from the projected data.
Further processing of the three-dimensional data to obtain the positions and volumes of the particles is performed by a segmentation algorithm using the DIPimage library \cite{hendriks1999dipimage} for matlab.}

\begin{figure}[]
\centering
  \includegraphics[width=8.6cm]{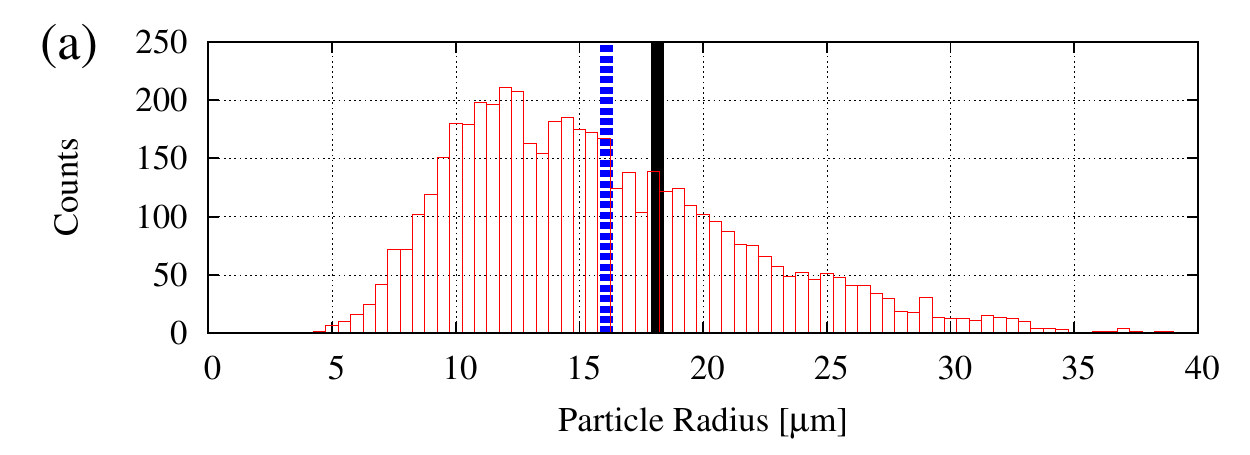}
  \includegraphics[width=8.6cm]{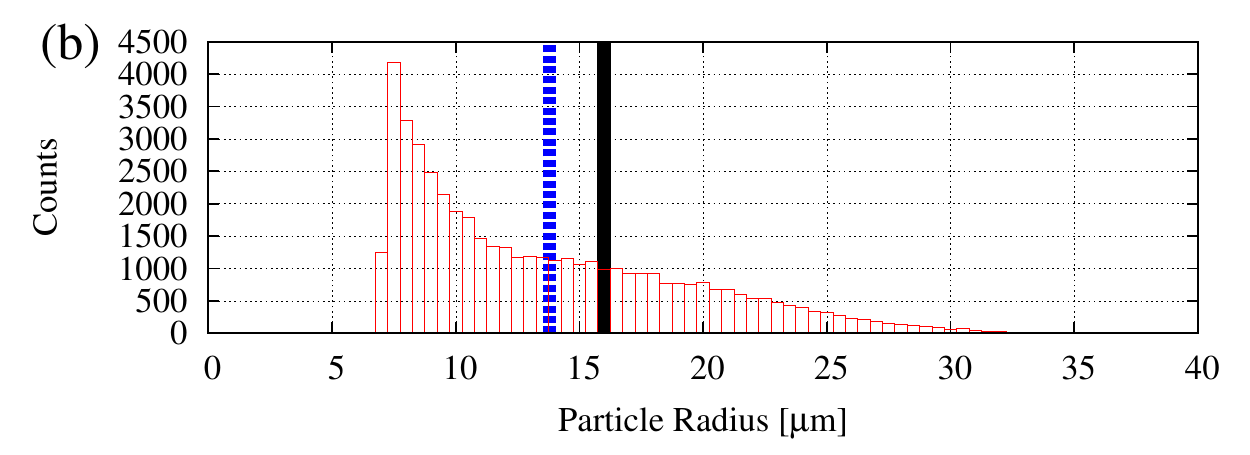}
  \caption{Distribution of radii in $\mu$m \GP{after converting the irregular shapes of the particles to spheres, maintaining their volumes as provided from the experimental analysis} for our (a) $15$ wt\% and (b) $40$ wt\% systems.
  Average radius $\sum_{i=1}^{N_p}a_i/N_p$ and unit length $\unitl=\sqrt[3]{ \sum_{i=1}^{N_p} a_i^3/N_p }$ are marked in the plots by the vertical dashed and solid lines, respectively.}
  \label{fig_radii_distr_iso40}
\end{figure}

\GP{The experimentally investigated particles had very different, irregular shapes.
To handle them effectively in our theoretical approach, we converted them to spheres of equivalent volume, which ensures that the overall magnetic dipole moment under saturated magnetization is maintained. 
The distribution of the resulting radii is shown in Fig.~\ref{fig_radii_distr_iso40} for the two systems.}
\GP{Moreover, the detection algorithm for our purpose had to be optimized for positional data, leading to a trade-off concerning the volume data, which does affect the size distributions for the two systems in Fig.~\ref{fig_radii_distr_iso40}.
The average radii and average cubed radii show some variations but the extracted radii basically stay within the range of $5-35\ \mu\mbox{m}$.
Both aspects hinder a quantitative comparison between experiments and theoretical results at the present stage, but good qualitative agreement is achieved.}

\begin{figure}[]
\centering
  \includegraphics[width=8.6cm]{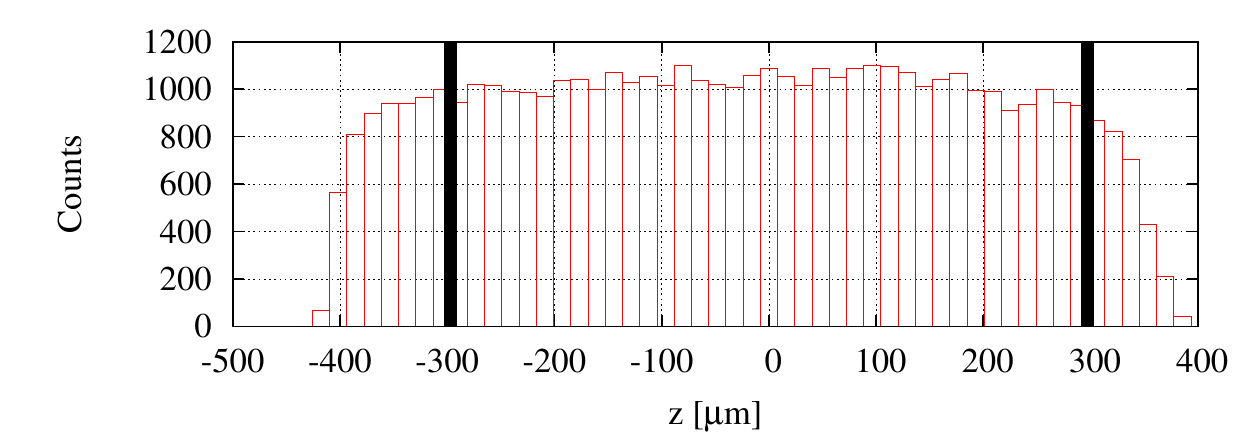}
  \caption{Distribution of the particles along the $z$-direction in $\mu\mbox{m}$ for the cylindrical sample of weight fraction $40$\%.
  To assure a uniform distribution of magnetic particles we numerically cut a cube from the center of the cylinder as indicated by the vertical lines.}
  \label{fig_z_distrib_iso40}
\end{figure}

\begin{figure}
\centering
  \includegraphics[width=8.6cm]{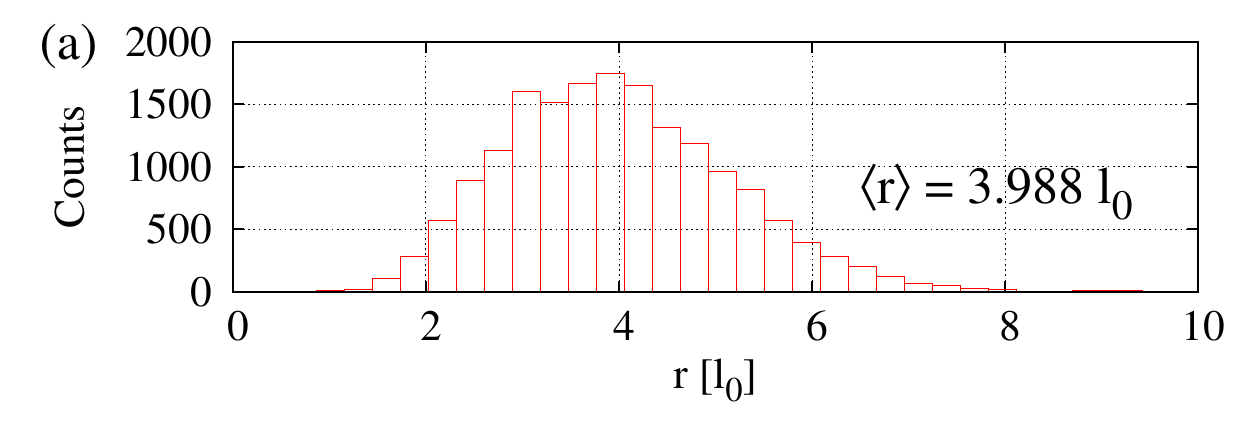}
  \includegraphics[width=8.6cm]{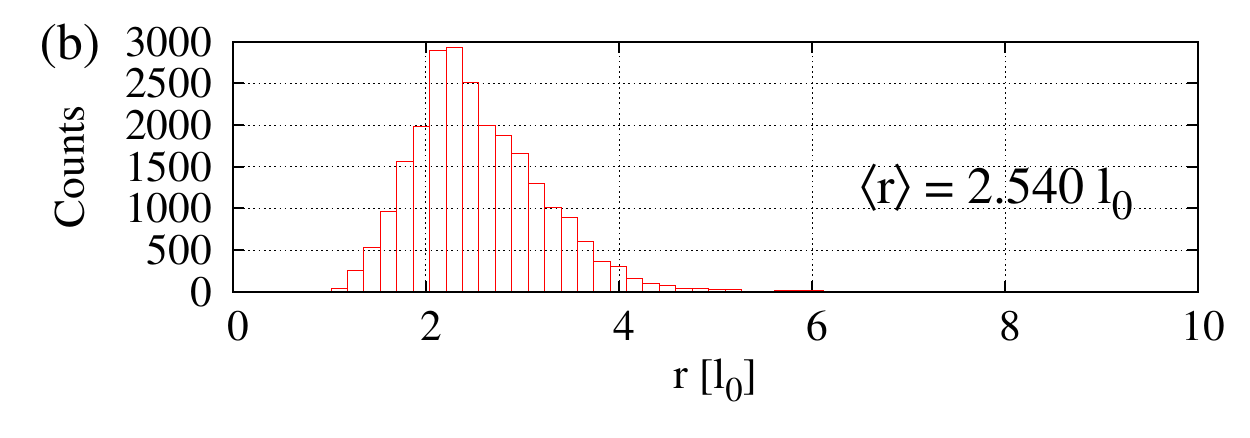}
  \caption{Distribution of the nearest-neighbor distances in our samples of (a) $15$ wt\% and (b) $40$ wt\% in terms of scaled particle distances $r/\unitl$.
  Particles on the boundaries are excluded from this calculation.}
  \label{fig_nn_distrib_iso}
\end{figure}

We choose the $z$-axis of our Cartesian coordinate system along the cylinder axis.
Then, we check the homogeneity of the samples by calculating the particle distribution along the $z$-direction, see Fig.~\ref{fig_z_distrib_iso40}.
As shown in Fig.~\ref{fig_z_distrib_iso40}, the particles are not completely uniformly distributed along the $z$-direction.
Variations are particularly ascribable to slight deviations from perfectly flat boundaries \cite{gundermann2017statistical}.
To work with a distribution of relatively uniform particle density, we use as input of our analysis the magnetic particles contained within a central cube of dimension $\simeq 600 \ \mu\mbox{m}$.
The distribution of mutual particle distances scaled by the length $\unitl$ is shown in Fig.~\ref{fig_nn_distrib_iso}.

Corresponding data of particle positions and volumes are imported into our numerical algorithm.
Then, in both cases a numerical mesh of initial edge length $a_{mesh}\simeq 2.5 \unitl$ is generated.
The total number of generated mesh nodes is much larger than the number of magnetic particles, see Tab.~\ref{Np_tab}.
\begin{table}
 \begin{center}
\renewcommand{\arraystretch}{1.0}
\setlength{\tabcolsep}{8pt}
  \begin{tabular}{ c  c  c  c c }
    \hline
    wt\% & $N_p$ & $N_{extra}$ & $N$ \\ \hline
    15\% & 186    & 4445   & \np \\
    40\% & \nmagn & \nmesh & \np \\
    \hline
  \end{tabular}
 \end{center}
 \caption{For each experimental sample of wt\% as indicated we select a cube from its center containing $N_p$ magnetic particles.
 $N_{extra}$ additional nodes are included in the spring network to achieve a total of $N = N_p + N_{extra}$ mesh nodes, each with $3$ translational degrees of freedom.}\label{Np_tab}
\end{table}
In the denser sample, we employ $N_{extra} =\nmesh$ extra nodes and $N_p = \nmagn$ magnetic particles, for a total of $3N=3\times\np = \ndof$ translational degrees of freedom, as indicated in Tab.~\ref{Np_tab}.
This amount of nodes corresponds to a total of $\Nsprings$ interconnecting springs. %spanning the whole sample.

\section{Calculation of Dynamic Moduli}\label{dyn_mod_calc_section}
In this section we briefly summarize our extended method to calculate the dynamic moduli.
We consider the system initially in equilibrium in the ground state for $M =|\bm{M}| =0$.
As $M$ increases, the particles reposition from the initial ground state to reduce their magnetic interaction energy $U^m$ until the closest minimum of the total energy $U$ is found, i.e.,
\begin{equation}\label{equilR}
 -\grad{U}{\vR_i} = \bm{F}_i =  \bm{0}, \ \ \ \ \forall \ i=1\dots N.
\end{equation}
This, in general, is achieved at the cost of increasing the elastic and steric energies $U^{el}$ and $U^s$.
Because of the large number of degrees of freedom and the irregular particle arrangement, the only practical way to find the equilibrium state is to perform numerical minimization.
In the present work, we chose to employ the FIRE algorithm \cite{bitzek2006structural}.
The control gauge for reaching the equilibrium state is the magnitude of the largest total force acting on a single particle.
This means that at convergence no particle is subject to a total force larger than a certain threshold, which we here set to $\threshold\unitforce$.

While obtaining the equilibrium configuration, it is crucial to suppress rigid translations and rotations of the whole system.
On the one hand, \GP{rigid translations induce} the system to drift in space and we tackle \GP{them} by subtracting an \GP{identical average} from all forces, i.e., by substituting
\begin{equation}
 \bm{F}_i \leftarrow \bm{F}_i -\frac1N \sum_{j=1}^N \bm{F}_j \ \ \forall i.
\end{equation}
Thus, the total force vanishes, $\sum_{i=1}^N \bm{F}_i =\bm{0}$.
On the other hand, a net overall rotation would alter the relative orientation of $\bm{M}$ with respect to the system boundaries.
Since we here intend to evaluate the moduli for specified geometries and orientations of $\bm{M}$, a global rotation of the system must be averted.
For this purpose, we first calculate the instantaneous overall torque $\bm{\tau}=\sum_{i=1}^N \bm{R}_i\times\bm{F}_i$.
As explained in section \ref{model_intro_section}, we model our system as overdamped.
Such a torque $\bm{\tau}$ would induce an instantaneous angular rotation $\bm{\omega}= \mathcal{I}^{-1}\cdot\bm{\tau}$ with the $3\times 3$ tensor $\mathcal{I}=c_0\sum_{i=1}^N\left(\bm{R}_i^2\,\mathds{1}-\bm{R}_i\bm{R}_i\right)$, its inverse $\mathcal{I}^{-1}$, and $\mathds{1}$ the unit matrix.
Thus, at every step of iteration, the force field acting on all the particles is rendered torque-free by subtracting from the force on each particle $i$ a counter-rotational component $\unitc \bm{\omega}\times\bm{R}_i$, i.e., by substituting
\begin{equation}
 \bm{F}_i \leftarrow \bm{F}_i -\unitc \left(\mathcal{I}^{-1} \cdot \sum_{j=1}^N \bm{R}_j \times \bm{F}_j \right)\times\bm{R}_i \ \ \forall i.
\end{equation}
Then, the total torque vanishes, $\bm{\tau}=\sum_{i=1}^N \bm{R}_i \times \bm{F}_i = \bm{0}$.
%Finally, we remark that these constraints do not need to be included into the total potential energy $U$ via Lagrange multipliers because global translations and rotations cost no energy.

Once the equilibrium positions $\vR_i^{\text{eq}}$ ($i=1\dots N$) are obtained under a given $\bm M$, we can calculate the corresponding dynamic moduli.
It is more convenient to treat the problem in terms of deviations from the equilibrium.
Therefore, we introduce the displacements $\vu_i=\vR_i-\vR_i^{\text{eq}}$ and their bra-ket notation $\left| \vu\right\rangle$ to indicate a $D$-dimensional (here $D=3N$) vector containing all the degrees of freedom.
The key object in this analysis is the Hessian matrix $\mathcal{H}$ of the total energy $U$.
Its elements are given by the second derivative of $U$ with respect of all degrees of freedom, $\mathcal{H}_{ij}=\partial_{u_i}\partial_{u_j}U$.
Since the system is in an energetic minimum, $U$ is a convex function of $\left| \vu\right\rangle$ and $\mathcal{H}$ is positive-semidefinite.
We denote eigenmodes and eigenvalues of $\mathcal{H}$ by $\left| \vv_n\right\rangle$ and $\lambda_n$, respectively, so that $\mathcal{H}\left| \vv_n\right\rangle=\lambda_n \left| \vv_n\right\rangle$.
If the system is subject to a small static external force field $\left| \vf \right\rangle$ acting on the mesh nodes, its final deformation is determined from the condition $\left| \vf\right\rangle = \mathcal{H}\left| \vu\right\rangle$.

\begin{figure}
\centering
  \includegraphics[width=8.6cm]{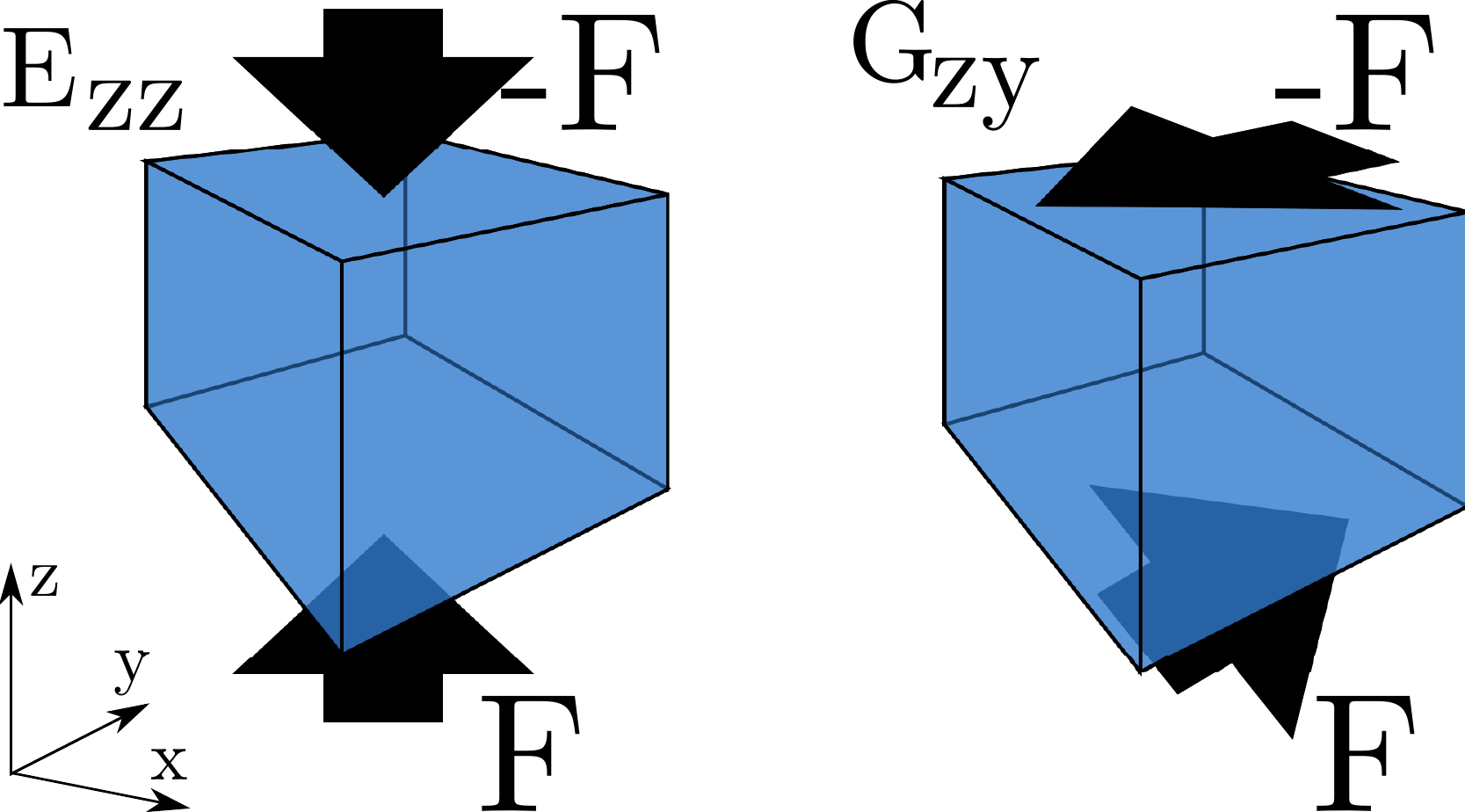}
  \caption{Main geometries to calculate the elastic Young ($E_{zz}$) and shear ($G_{zy}$) moduli.}
  \label{el_geom}
\end{figure}

To describe external force fields that result in axial or shear deformations, %\cite{pessot2016dynamic}, 
we define external forces acting on the boundary particles.
They represent a mechanical stress applied from outside and oriented in a preselected direction.
In the present work, sticking to the experimental set-up, we focus on axial stretching/compression along the $z$-direction and shear strains with the shear plane containing $\widehat{\bm{z}}$ but the force applied perpendicular to it, see Fig.~\ref{el_geom}.
A corresponding force field must satisfy the following conditions: i) it is non-vanishing only on the boundaries and oriented in the appropriate directions to build up the corresponding macroscopic stress; ii) it induces neither translational drift iii) nor net rotations; and iv) its magnitude scales with the macroscopic force $F$ corresponding to the stress $\sigma=F/S$ acting on the boundary of surface $S$.

When a small external force is applied, the system rearranges to balance it and its total energy increases.
By decomposing the external force field over the eigenmodes of deformation \cite{tarama2014tunable, huang2015buckling, babel2016dynamics, pessot2016dynamic}, we can identify the contribution of each mode to the total change in energy.
The corresponding static elastic modulus is found as
\begin{equation}\label{static_elm}
 E = \frac{L}{S} {\left[ \frac{1}{F^2}\sum_{n=1}^D \frac{ {\langle \bm{f} | \bm{v}_n \rangle}^2 }{\lambda_n} \right]}^{-1},
\end{equation}
where $L$ is the distance between the boundaries on which the forces are applied.
It is calculated later as described in section \ref{results_strain_section}.
Moreover, the surface $S$ of the boundary $\mathcal{B}_{\widehat{\bm{z}}}$ is obtained by projecting the nodes and particles belonging to it onto the plane perpendicular to $\widehat{\bm{z}}$.
Subsequently, the convex hull of the projected set of points is determined \cite{jones2001scipy}, which allows us to describe samples of non-regular and changing shape. % \cite{pessot2016dynamic}.

In the case of a periodically oscillating stress, the former static condition $\left| \vf\right\rangle = \mathcal{H}\left| \vu\right\rangle$ becomes an overdamped equation of motion which reads
\begin{equation}\label{eq_motion}
 \mathcal{C} \left|\dot{\bm{u}}\right\rangle(t) + \mathcal{H} |\vu\rangle (t) = |\vf\rangle (t).
\end{equation}
The first step consists in transforming it into an eigenvalue problem.
The entries of the matrix $\mathcal{C}$ of viscous friction read $\mathcal{C}_{ij}= \unitc \delta_{ij}$, with $\delta_{ij}$ the Kronecker delta and $i,j=1\dots N$.
Since $\mathcal{C}$ and $\mathcal{H}$ commute and can be diagonalized simultaneously, Eq.~(\ref{eq_motion}) decouples into $D=3 N$ one-variable equations
\begin{equation}
 \unitc \dot{u}_n(t) +\lambda_n u_n(t) = f_n (t)
\end{equation}
($n=1\dots 3 N$).
Each describes the dynamics of a single normal mode, with ${u}_n(t)= \langle \bm{v}_n  \left|{\bm{u}}\right\rangle(t)$ and ${f}_n(t)= \langle \bm{v}_n  \left|{\bm{f}}\right\rangle(t)$.

If the time-dependent external force is periodic with a single frequency, i.e., $|\vf\rangle (t)=|\vf^0\rangle \e^{\imu \omega t}$, then the solution $|\vu\rangle (t)= |\vu^0\rangle \e^{\imu \omega t}$ of Eq.~(\ref{eq_motion}) in the steady state will be oscillating with identical frequency, possibly with a time lag.
The same is true for the projections onto the normal modes ${f}_n(t)$ and ${u}_n(t)$, respectively.
Solving for each normal mode in the steady state, we find
\begin{align}\label{linr_func}
u_n(t) = u^0_n \e^{\imu \omega t} &= \frac{f^0_n \e^{\imu \omega t}}{\kappa_n(\omega)} = \frac{f_n(t)}{\kappa_n(\omega)}
\end{align}
with $\kappa_n(\omega)= \lambda_n +\imu \unitc \omega$, $u_n^0=\langle\vu^0|\vv_n\rangle$, and $f_n^0=\langle\vf^0|\vv_n\rangle$ \cite{pessot2016dynamic}.
Thus, starting from a given external oscillating force field $|\vf^0\rangle \e^{\imu \omega t}$ we can calculate the dynamic linear response of the system in the form $|\vu^0\rangle \e^{\imu \omega t}$.

Finally, we define the complex single-frequency dynamic elastic moduli as the ratio between stress and strain in the steady state regime: $E(\omega)=E'(\omega) +\imu E''(\omega)= \sigma(t)/ \varepsilon(t)$.
The real and imaginary parts $E'(\omega)$ and $E''(\omega)$ are defined as storage and loss moduli, respectively.
To bridge the gap between the macroscopic and the mesoscopic quantities, we identify the total strain of the system as the displacement of the forced boundary particles over the distance between the forced boundaries, i.e., $\varepsilon(t)=\langle\vu(t)|\vf^u\rangle/L$.
Here, $|\vf^u\rangle$ indicates a force field rescaled so that it exerts a total force of unitary magnitude on each boundary in the overall force direction.

The dynamic elastic moduli are then calculated as \cite{pessot2016dynamic}
\begin{equation}\label{dyn_elm}
 E(\omega) = \frac{L}{S} {\left[\sum_{n=1}^D \frac{ {f_n^u}^2 }{\kappa_n(\omega)} \right]}^{-1},
\end{equation}
with $f^u_n=\langle\vv_n|\vf^u\rangle$.

To summarize, we have outlined a procedure that from mesoscopic particle distributions and discretized mesoscopic force fields yields the macroscopic stresses, strains, and elastic moduli.
In the following, we first characterize the particle distributions obtained from the experiments by volume fraction of magnetic particles, particle radii, extension along the $z$-coordinate, and distribution of nearest-neighbor distances.
Then, we consider the effect of increasing magnetic interactions.
Our main focus will be on the resulting changes in length and dynamic elastic moduli.

\section{Results}\label{results_section}
We now investigate how increasing magnetization $M=|\bm{M}|$ of the particles affects the mechanical properties of the system such as axial strain and elastic moduli.
In agreement with the experimental set-up, %see section \ref{micro_tomo_section}, 
we here set $\bm{M} =M\widehat{\bm{z}}$.
For the presentation of our results, we measure lengths, energies, forces, and elastic moduli in multiples of $\unitl$, $\uniten$, $\unitforce$, and $\unitk$, respectively.
Viscosity, velocities, times, and frequencies are measured, respectively, in multiples of $\unitc/\unitl$, $\unitk{\unitl}^2/c_0$, $c_0/\unitk{\unitl}$, and $\unitfreq=\unitk{\unitl}/c_0$.
Finally, we measure magnetic moments and magnetization in multiples of $\unitmom=\sqrt{4\pi\unitk \unitl^6/\mu_0}$ and $\unitmag=\sqrt{4\pi\unitk/\mu_0}$, respectively.

Since $\unitk$ scales the elastic moduli of the matrix, $\unitmag=\sqrt{4\pi\unitk/\mu_0}$ gauges the relative strength of elastic and magnetic effects in our reduced units.
A magnetoelastomer with elastic modulus of $\sim 10^4 \textrm{ Pa}$ \cite{varga2006magnetic, schuemann2017insitu} implies $\unitmag$ of the order of $\sim 3\times10^5 \textrm{ A/m}$.
Since the saturation magnetization of carbonyl iron is $\sim 2\times10^6 \textrm{ A/m}$ \cite{bombard2003magnetic}, the range of magnetization would be $M \lesssim 7 \unitmag$.
However, applying the rescaled model to soft gels, implying $\unitk \sim 1 \textrm{ Pa}$ \cite{huang2015buckling}, suggests up to $M \lesssim 30\unitmag$ in reduced units.

We consider deformations of the system explicitly involving the magnetization orientation, i.e., the $z$-direction. Accordingly, we focus on the elastic moduli $E_{zz}$ and $G_{zy}$ corresponding to axial strain in the $z$-direction and shear strain with the shear plane containing $\bm{M}$, respectively, as depicted in Fig.~\ref{el_geom}.

\subsection{Field-induced Strain and Chain Formation}\label{results_strain_section}
In the following, we indicate with the term ``magnetostriction'' (sometimes also referred to as ``magnetodipolar striction''\cite{stolbov2011modelling}) the change in length of the sample under the effect of increasing magnetization.
In experiments, magnetoelastic materials can show elongation \cite{szabo1998shape, diguet2010shape} or contraction \cite{zhou2004deformation, coquelle2005magnetostriction} along the field direction.
As pointed out by some authors, the magnetostrictive response can qualitatively change according to the underlying particle distribution \cite{stolbov2011modelling}, shape of the sample \cite{diguet2010shape}, volume fraction of magnetic particles, as well as magnetic field intensity \cite{zubarev2012theory}.

We are mainly interested in length changes along the $z$-axis.
For this purpose, we define the length in the $z$-direction as $L_z= {\langle z \rangle}_{+\widehat{\bm{z}}} -{\langle z \rangle}_{-\widehat{\bm{z}}}$, where ${\langle\dots\rangle}_{\pm\widehat{\bm{z}}}$ indicates averages over the subsets of nodes belonging to the $\pm\widehat{\bm{z}}$ boundaries, respectively.

\begin{figure}
\centering
  \includegraphics[width=8.6cm]{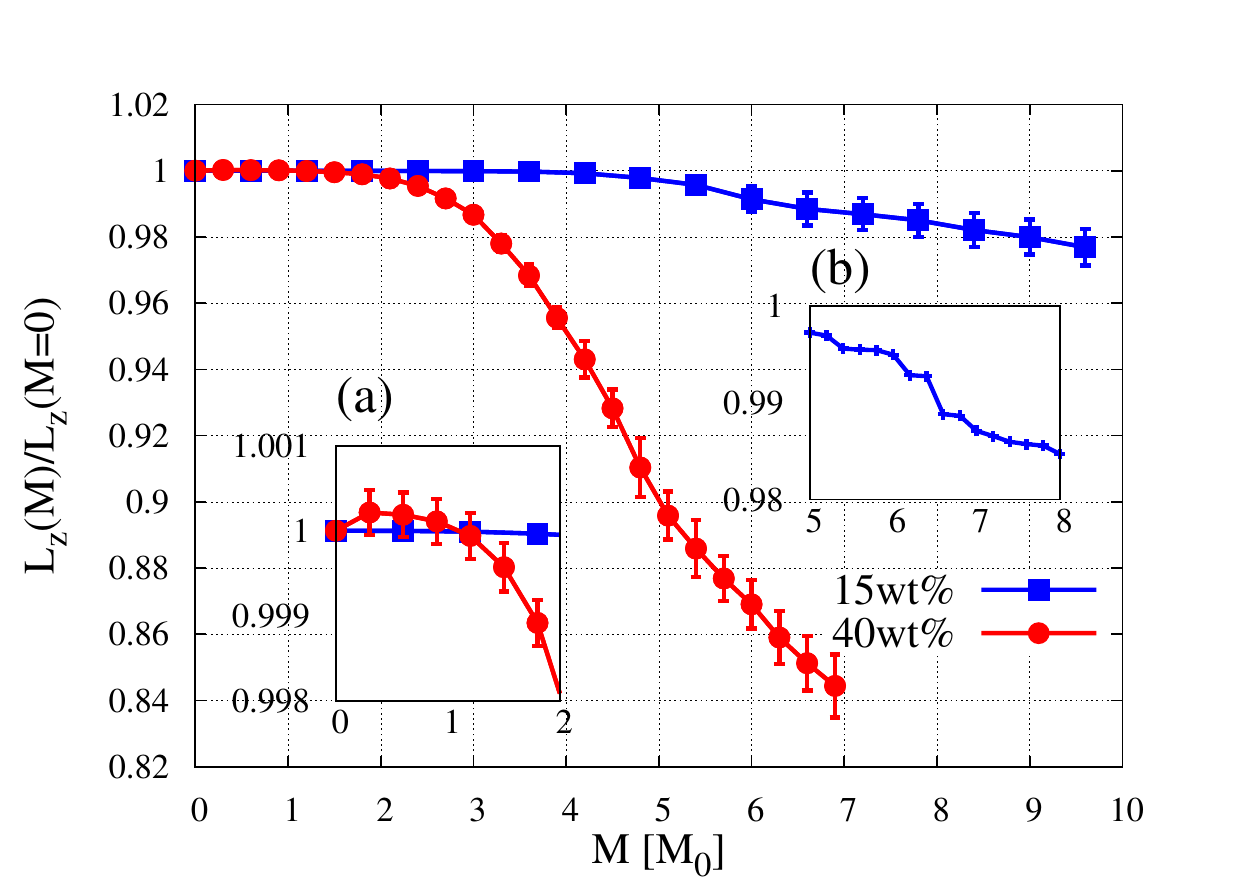}
  \caption{Changes in distance $L_z$ between top and bottom boundaries for increasing magnetization $M$ ($\bm{M}\parallel\widehat{z}$) and varying particle concentration ($15$ and $40$ wt\%).
  Data points and bars represent, respectively, averages and standard deviations over $\Nsamples$ differently randomized spring networks, while the positions of the magnetic particles are unchanged.
  Inset (a) zooms in onto the initial behaviors for small $M$.
  Inset (b) depicts a zoom-in of an individual simulation, in which discontinuities caused by the formation of chains are visible.}
  \label{z_strain_iso}
\end{figure}

The change in length $L_z$ for the two samples of varying wt\% with increasing $M$ are depicted in Fig.~\ref{z_strain_iso}.
As a dominant trend, we observe contraction in the direction of magnetization.
This follows from the overall attractive role played by the dipole--dipole interaction in the systems.
However, at small values of $M$, a small initial elongation is observed for the $40$ wt\% sample, see inset (a) of Fig.~\ref{z_strain_iso}.
Such an initial elongation appears due to the initial repositioning of the particles.
The initial tendency of dipoles to rotate around each other to minimize the magnetic energy, e.g., as in the fcc case described in Ref.~\onlinecite{pessot2016dynamic}, could be responsible for such an initial elongation.
For larger $M$, however, the dipole--dipole attraction in the $\bm{M}$-direction %is much stronger and 
induces contraction.
Apart from that, it is evident from Fig.~\ref{z_strain_iso} that higher volume fractions of magnetic particles increase the deformational response, as one would expect intuitively.
Naturally, higher volume fractions imply smaller interparticle distances, and therefore stronger dipole--dipole interactions, see Eq.~(\ref{emagn}).

\begin{figure}
\centering
  \includegraphics[width=8.6cm]{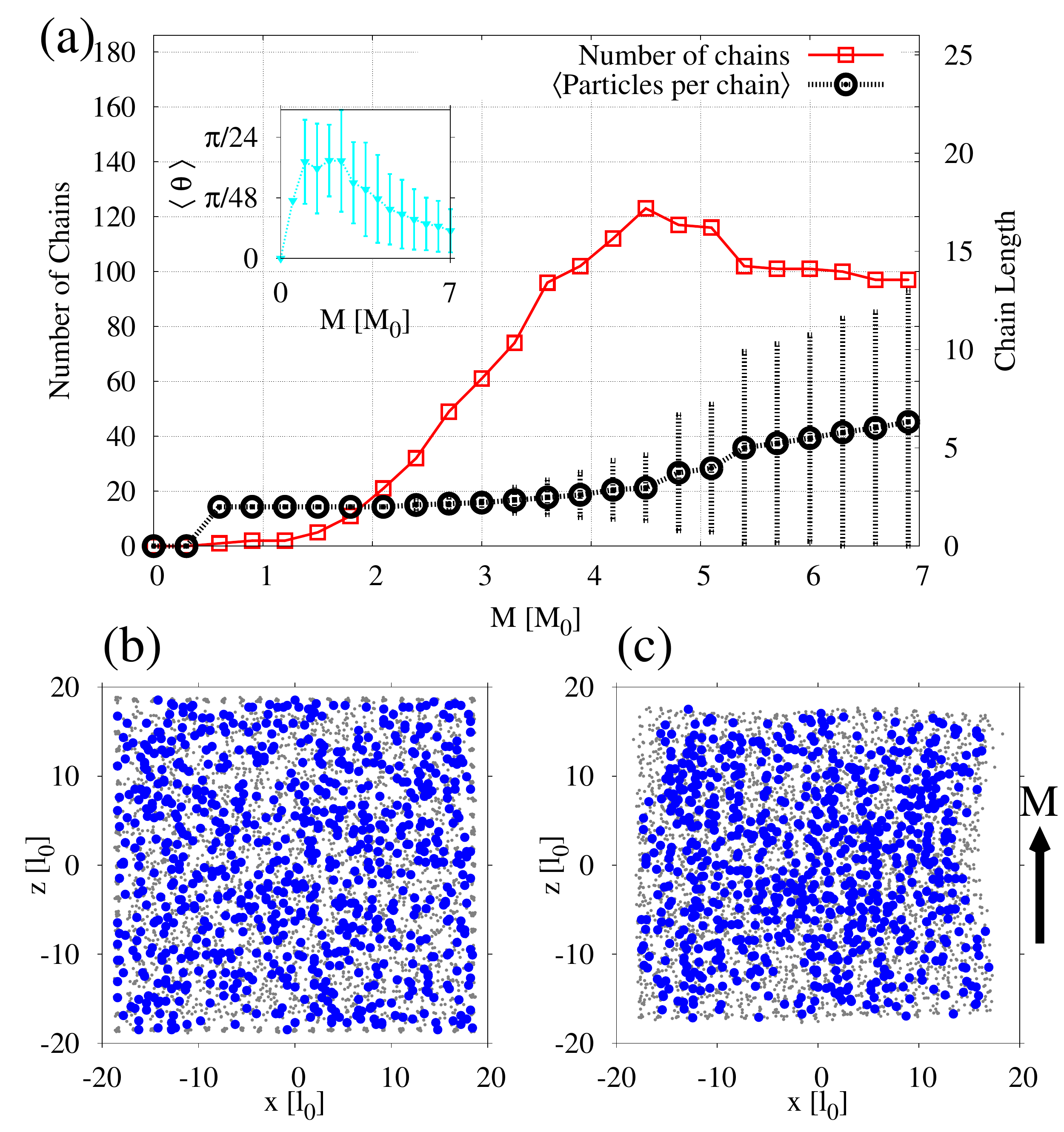}
  \caption{Clustering in one numerical discretization of the $40$ wt\% sample.
  Panel (a) shows the number of chains formed for increasing $M$ (left ordinate axis), together with the average number of particles comprised by each chain (right ordinate axis) plotted by solid and dotted lines, respectively.
  Furthermore, points and bars represent averages and standard deviations over the set of aggregates present in the system.
  The inset addresses the angle $\theta$ between the end-to-end vectors of the chains and $\bm{M}$ for increasing $M$.
  Panel (b) depicts the projection of nodes (light-colored) and magnetic particles (dark-colored) at $M=0$ onto the $xz$ plane.
  Panel (c) shows the same plot for the largest $M$ reached in panel (a).
  There, the formation of chain-like aggregates as well as the overall deformation of the spring network are visible.}
  \label{iso40_clustering}
\end{figure}

We remark that the deformation of the sample is approximately continuous only in the small-$M$ regime.
For larger values of $M$, jumps in the degree of deformation are observed for the individual systems, see inset (b) of Fig.~\ref{z_strain_iso}.
This results from the granular mesoscopic resolution of the magnetic particles and occurs when the dipole--dipole attraction between two magnetic particles is strong enough to overcome the linear springs connecting them.
Then, in a similar fashion as described in Ref.~\onlinecite{annunziata2013hardening} but in a $3$D environment, the particles collapse towards each other along the $\bm{M}$-direction and are then stabilized by steric repulsion.
Each such process strongly contributes to the contraction along $\bm{M}$.
Consequently, chain-like clusters start to grow which, initially, consist of just two particles, as depicted in Fig~\ref{iso40_clustering}.
For $M \gtrsim 2.5 \unitmag$ chains of more than $2$ particles begin to form.
After a large enough amount of aggregates has formed, the chains start to merge with each other.
This is \GP{signaled by a decreasing} number of chains for $M \gtrsim 4 \unitmag$ while the average chain length keeps increasing.

\subsection{Dynamic Elastic Moduli}\label{results_dynmods_section}
\begin{figure}
\centering
  \includegraphics[width=8.6cm]{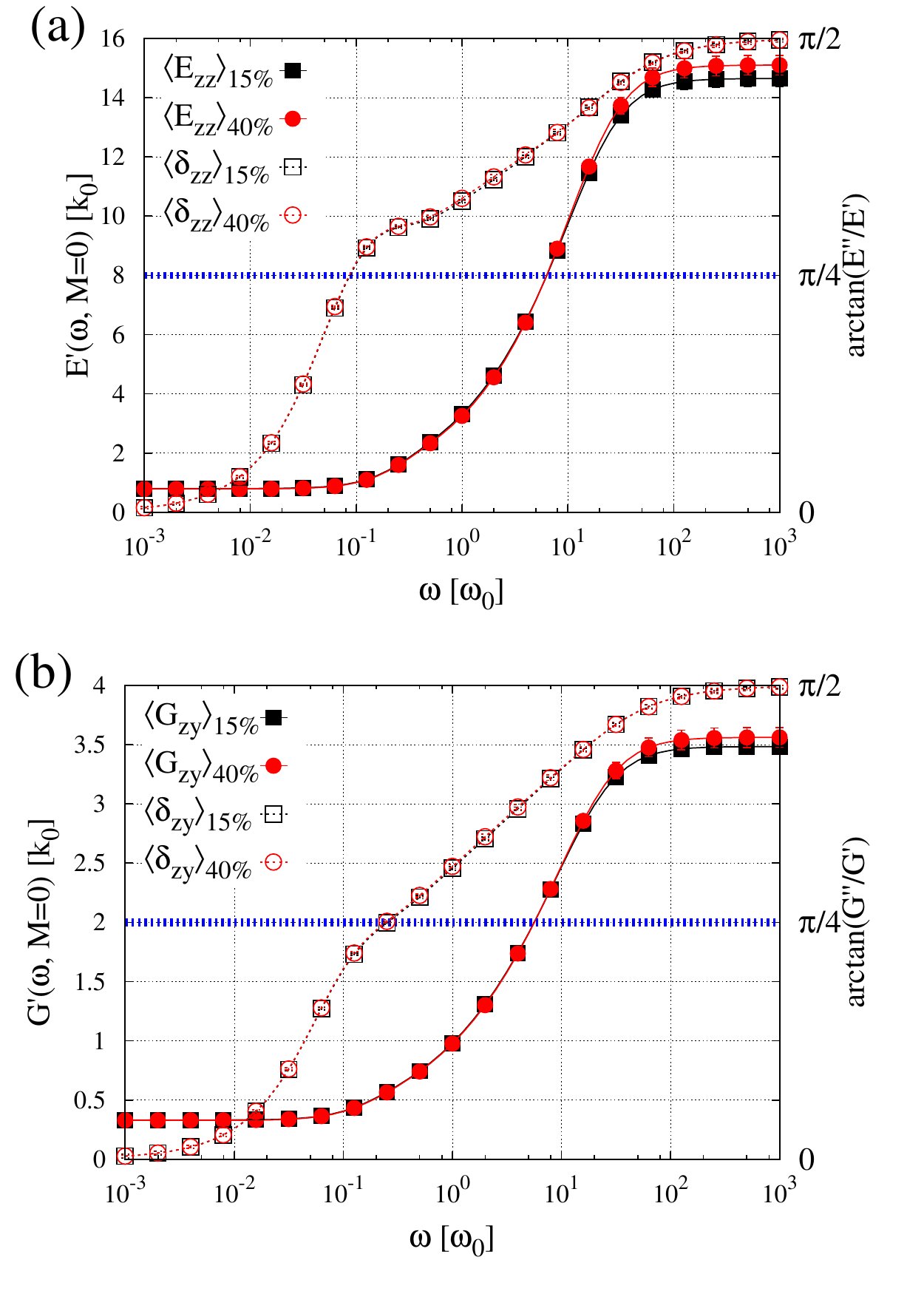}
  \caption{Dynamic Young and shear moduli $E_{zz}(\omega)$ and $G_{zy}(\omega)$ for vanishing magnetic interactions ($M=0$) for arrangements extracted from samples of $15$ and $40$ wt\%, respectively.
  The curves for the storage moduli $E'_{zz}(\omega)$ and $G'_{zy}(\omega)$ refer to the left axes.
  The curves for the phase lags between storage and loss components, $\delta_{zz}(\omega)=\arctan(E''_{zz}/E'_{zz})$ and  $\delta_{zy}(\omega)=\arctan(G''_{zy}/G'_{zy})$, refer to the right axes.
  For $\delta_{zz}(\omega) = \pi/4 = \delta_{zy}(\omega)$ (horizontal dashed blue line) corresponding storage and loss moduli are equal.
  Data points and bars (where visible) represent averages and standard deviations over $\Nsamples$ different numerical realizations of the spring network surrounding the magnetic particles.}
  \label{dynmods_15_40}
\end{figure}

We now move on to the dynamic Young and shear moduli $E_{zz}(\omega)=E'_{zz}(\omega) +\imu E''_{zz}(\omega)$ and $G_{zy}(\omega)=G'_{zy}(\omega) +\imu G''_{zy}(\omega)$ as a function of frequency $\omega$ and particle magnetization $M$.
Estimating our reduced unit of measure for the frequency $\unitfreq=\unitk \unitl / \unitc$ requires knowledge of the friction coefficient $\unitc$. % relaxation rate of a mesoscopic particle moving in the polymer matrix.
\GP{Here, we} choose a different approach.
To compare with experimental data, we match the frequency at which the storage and loss moduli cross as shown in Fig.~\ref{dynmods_15_40}.
At this frequency, $\arctan(E'/E'')=\pi/4$.
Comparing with representative rheological measurements \cite{zanna2002influence}, we estimate that $\unitfreq \sim {10}^3 \mbox{ Hz}$ for a typical polymeric material of modulus $\sim 10^5 \mbox{ Pa}$.
Experiments in the low strain regime \cite{zanna2002influence, roth2010viscoelastic} can explore wide frequency intervals (${10}^{-1}-{10}^{6}\mbox{ Hz}$) that in our reduced units would correspond to ${10}^{-4}-{10}^{3}$~$\unitfreq$.

As explained in section \ref{dyn_mod_calc_section}, the dynamic moduli, e.g., $E_{zz}(\omega)$, link the macroscopic oscillating stresses and strains, i.e.,
\begin{equation}
 \sigma_{zz}(\omega)= E_{zz}(\omega) \varepsilon_{zz}(\omega) \label{dynamic_stress_strain}.
\end{equation}
In the steady-state regime stress and strain both oscillate with the same frequency but shifted by a phase 
\begin{equation}
 \delta_{zz}(\omega)=\arctan\left[\frac{E''_{zz}(\omega)}{E'_{zz}(\omega)}\right].\label{phase_lag}
\end{equation}

Our modeling of the particle dynamics as in Eq.~(\ref{eq_motion}) corresponds to a Kelvin--Voigt macroscopic model.
Such a model is particularly appropriate at longer timescales (i.e., in the small-$\omega$ regime).
There, it is characterized by a constant storage modulus and a loss modulus that linearly increases with the frequency.
Upon decomposition into the normal modes, each mode behaves as an independent Kelvin--Voigt element with different parameters and dynamic modulus $\kappa_n(\omega)=\lambda_n +\imu \unitc \omega$ as in Eq.~(\ref{linr_func}).
Increasing the oscillation frequency, the dynamic moduli deviate from a simple Kelvin--Voigt description, as shown in Fig.~\ref{dynmods_15_40}, because the response of the system switches to different combinations of modes.

Data points and bars in Fig.~\ref{dynmods_15_40} are obtained from averages and standard deviations of $\Nsamples$ uncorrelated, differently randomized numerical realizations of the spring network.
Furthermore, and albeit the Kelvin--Voigt model describes particularly the long-timescale behavior, we here plot for completeness a larger range of $\omega$.
We can mainly distinguish between three regimes of frequency.

First, up to $\sim 10^{-2}\omega_0$, the storage moduli have a flat behavior and, in the $\omega\rightarrow 0$ limit, \GP{converge to} the static elastic moduli.
Here the deformation occurs over long timescales and the bulk relaxes completely.
Therefore, the storage modulus has its minimum with respect to $\omega$.
Typically, in experiments the elastic moduli increase with increasing volume fraction $\phi$ of hard inclusions.
According to Einstein's law, $E(\phi) = E(\phi=0) [1+5\phi/2]$ to lowest order in $\phi$ \cite{einstein1906eine, smallwood1944limiting}.
Here we do not observe this effect, because we set the springs between the centers of the particles.

\begin{figure}
\centering
  \includegraphics[width=8.6cm]{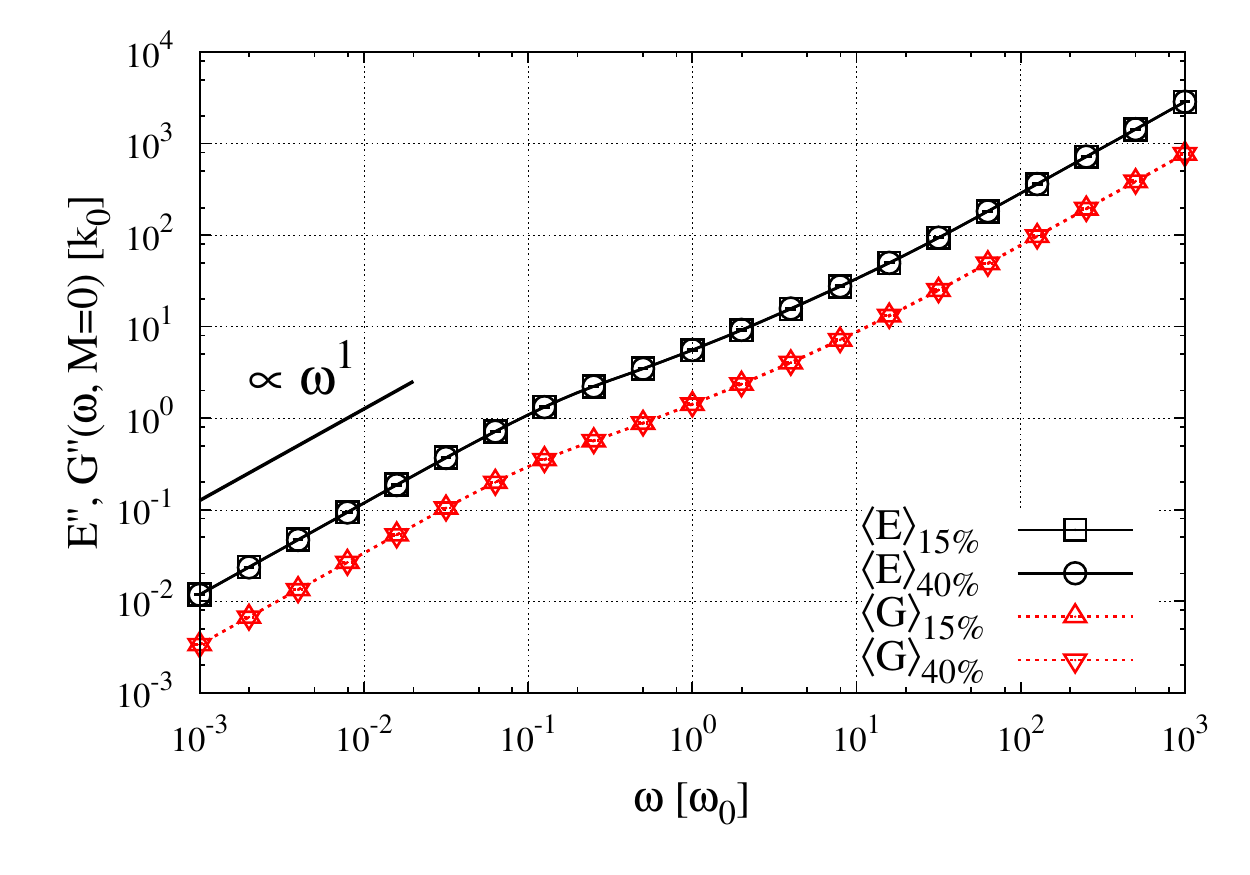}
  \caption{Loss moduli $E''_{zz}(\omega)$ (solid line) and $G''_{zy}(\omega)$ (dashed line) at vanishing magnetization $M=0$ for samples with $15$ and $40$ wt\%.
  The double logarithmic scale reveals the initial linear behavior.
  Data points indicate averages over $\Nsamples$ different numerical realizations of the spring network surrounding the magnetic particles.
  Standard deviations are much smaller than the symbols.}
  \label{lossmods_15_40}
\end{figure}

The loss moduli in the low-$\omega$ regime linearly increase and follow the trend $\sim \eta \omega$, with $\eta$ an effective viscosity, see Fig.~\ref{lossmods_15_40}.
The viscosities are obtained from the initial slopes of $E''_{zz}(\omega)$ and $G''_{zy}(\omega)$.
They are, respectively, $E''_{zz}(\omega)/\omega\sim 10 \unitc/\unitl$ and $G''_{zy}(\omega)/\omega\sim 2 \unitc/\unitl$.
In this regime, the phase lags are increasing but small, see Fig.~\ref{dynmods_15_40}, so that stress and strain are almost completely in phase.

Increasing $\omega$ until $\sim 10^{-1}\omega_0$, the phase lags reach the value $\pi/4$, as a consequence of the increasing loss moduli.
At this characteristic frequency the storage moduli equal the corresponding loss moduli.
As indicated above, this reference point could be used to compare our results with experimental measurements.

For increasing frequency $\omega$, the bulk of the system is unable to relax as the oscillation period of the external stress decreases.
As a consequence we find an increase in the storage moduli up to $\omega \sim 10^{2}\omega_0$, where they reach a final plateau.
Here, the linear response results from the springs on the boundary and practically no dynamical internal relaxation occurs.
The frequency is too large to allow for propagation of the external stimulus into the bulk.
In agreement with the Kelvin--Voigt model, the loss moduli keep increasing.
Some deviations from the linear increase are visible in the regime $10^{-1}\omega_0 \lesssim \omega \lesssim 10^{1}\omega_0$, in which the system switches from a low-$\omega$ bulk to a high-$\omega$ surface response.
Subsequently, in the range $\omega \gtrsim 10^{2}\omega_0$, the phase lags practically reach the stationary value of $\pi/2$, for which stress and strain are completely out of phase.
We mention that, at high frequencies, inertial effects may become important.
In this case, our results obtained from overdamped dynamics may lose their significance.

\subsection{Hardening Effects}\label{results_harden_section}
Increasing the magnetization $M$ of the particles, e.g., by applying an external magnetic field, their spatial arrangement undergoes significant transformations, see Figs.~\ref{z_strain_iso}, \ref{iso40_clustering}, and section \ref{results_strain_section}.
Such adjustments are reflected by variations in the elastic moduli.

First we address induced changes in the static moduli $E_{zz}(M)$ and $G_{zy}(M)$, where $E_{zz}(M)=E'_{zz}(\omega=0,M)$ and $G_{zy}(M)=G'_{zy}(\omega=0,M)$.
For briefness, we denote the moduli at vanishing frequency and magnetization as $E_{0}=E'_{zz}(\omega=0,M=0)$ and $G_{0}=G'_{zy}(\omega=0,M=0)$.
Moreover, we focus our analysis on the sample with $40$ wt\% because it shows the stronger response to magnetic interactions.
Again we average our results over $\Nsamples$ different numerical realizations of the spring network and show the corresponding averages and standard deviations in our results.
In this way, we link our findings to the specific particle distribution and not to the specific arrangement of the network nodes.

\begin{figure}
\centering
  \includegraphics[width=8.6cm]{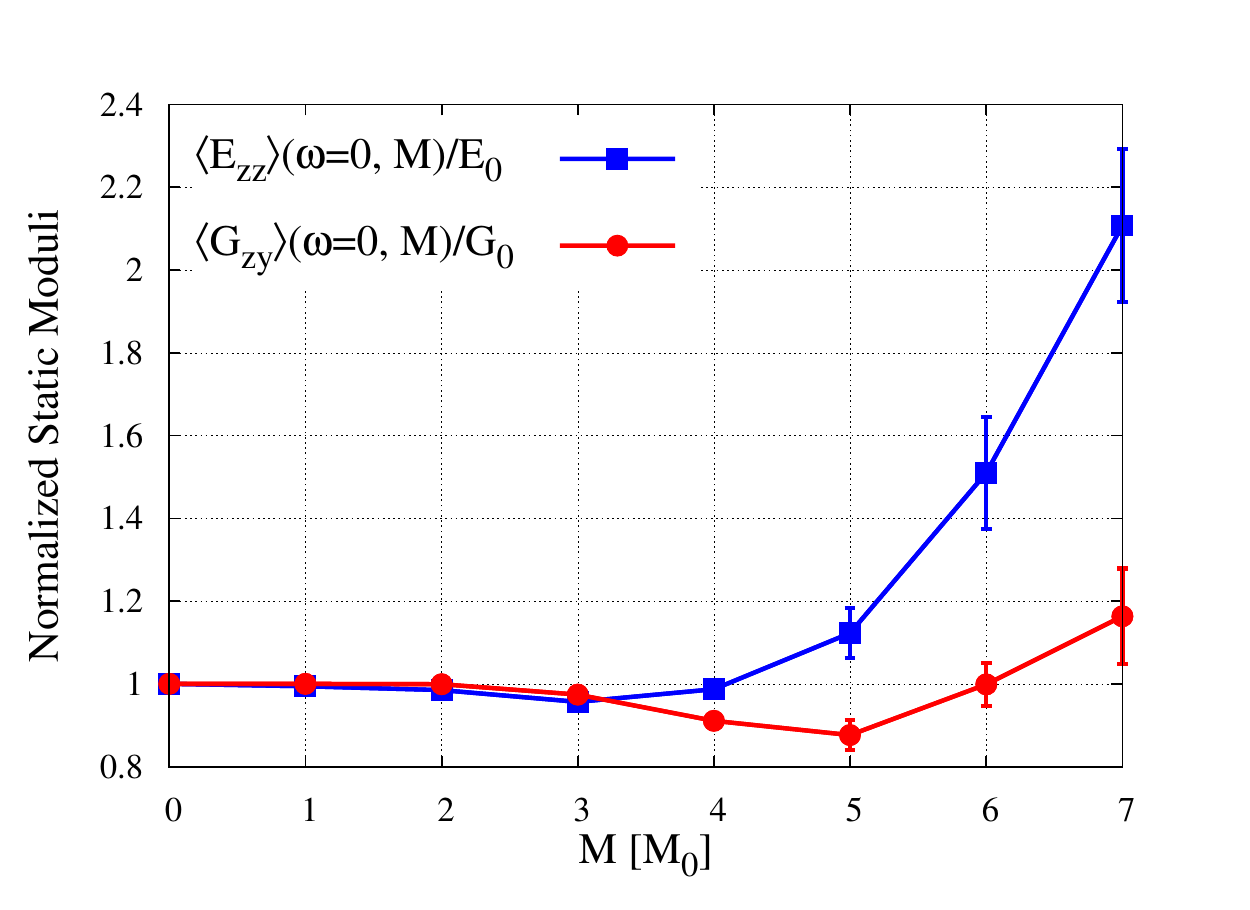}
  \caption{Variation of the static ($\omega=0$) normalized moduli $E_{zz}(M)/E_0$ and $G_{zy}(M)/G_0$ with increasing magnetization for the sample of $40$ wt\%.
  Data points indicate averages over $\Nsamples$ different numerical realizations of the spring network.
  For $M<5\unitmag$ the standard deviations are smaller than those at $M=5\unitmag$.}
  \label{iso_static_m}
\end{figure}

Initially, for $M\lesssim 2 \unitmag$ the moduli do not vary significantly, see Fig.~\ref{iso_static_m}.
In this low-$M$ regime, the amount of formed chain-like aggregates is quite small and only few particles are clustered.
Furthermore, here the standard deviations on the moduli due to the different realizations of the numerical spring network are less than $1$\%.

For values of $M\gtrsim 2 \unitmag$ we first observe a slightly decreasing trend for both moduli, more accentuated for the shear modulus which is reduced by up to $12\%$.
This behavior corresponds to a softening of the systems.
Significant particle rearrangements occur in this regime due to the induced magnetic interactions.
Apparently, the resulting energetic locking of the corresponding intermediate structures is weaker than in the initial unmagnetized state.
Such configurations are, however, stable minima of the system for each given $M$, as evidenced by the positive values of the moduli.

Finally, at higher $M$, we observe a significant increase in both moduli, together with an increase in the statistical standard deviations.
The corresponding large increase of elastic moduli has previously been predicted theoretically \cite{pessot2016dynamic} and been observed in experiments \cite{varga2006magnetic, kallio2007dynamic, schuemann2017insitu}.
We directly attribute this hardening of the sample to the formation of large chain-like aggregates in the system \cite{annunziata2013hardening}.
They are aligned in the $\bm{M}$-direction and can span large portions of the sample.
When the magnetic particles on the chains are at contact in a ``hardened'' \cite{annunziata2013hardening} state, they are virtually locked in position by the intense, counterbalancing steric and magnetic forces.
Each such particle is trapped in a potential well much steeper than the one originating from the spring network.
Thus, it is intuitive that the macroscopic deformations illustrated in Fig.~\ref{el_geom} experience a significant stronger resistance if they work on the hardened chains.
Moreover, it is intuitive that the increase in $E_{zz}$ is stronger than the one in $G_{zy}$ in Fig.~\ref{iso_static_m}

\begin{figure}
\centering
  \includegraphics[width=8.6cm]{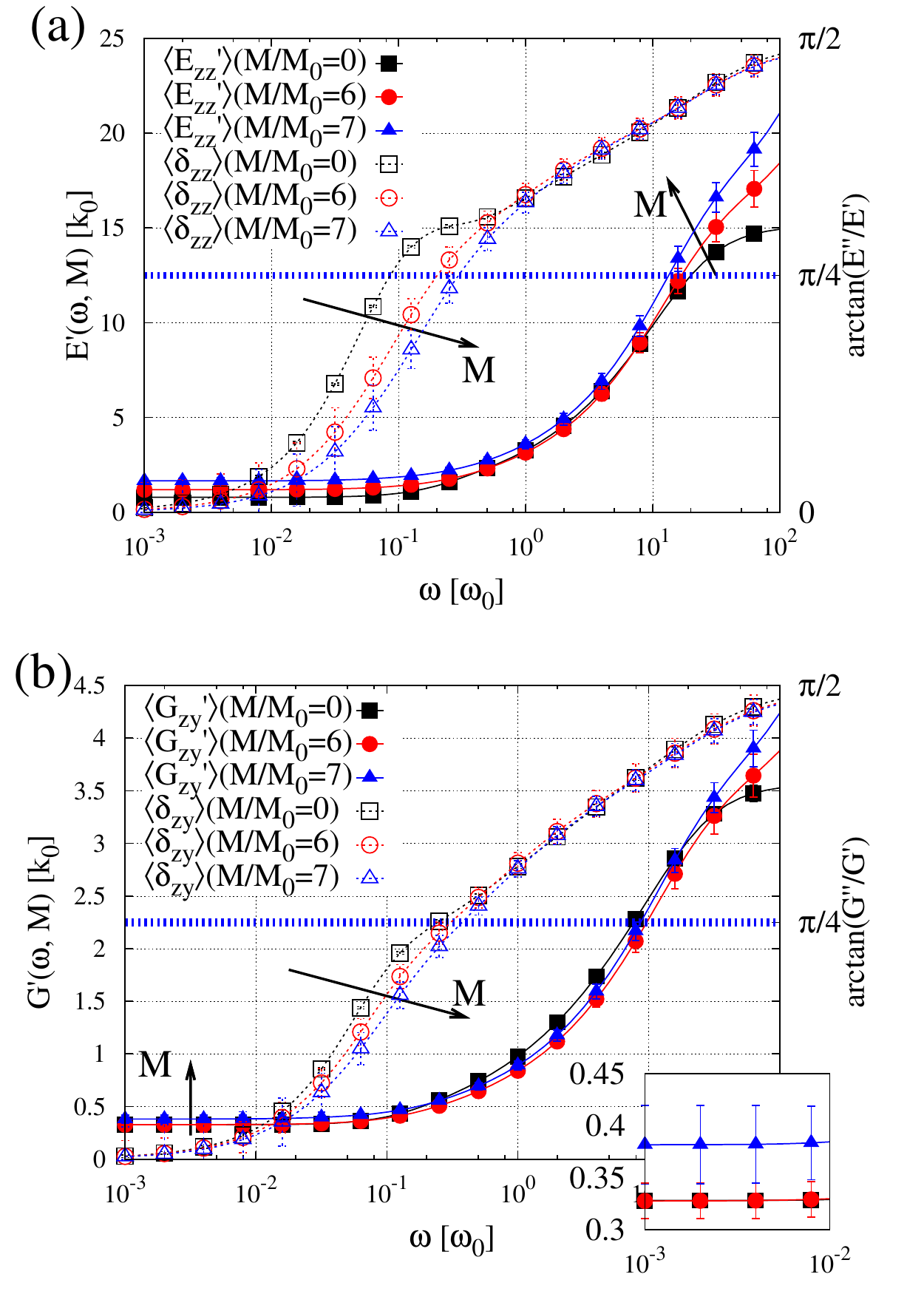}
  \caption{Effect of increasing magnetization $M$ on the storage moduli (a) $E'_{zz}(\omega)$ and (b) $G'_{zy}(\omega)$ as well as on the phase lags (a) $\delta_{zz}(\omega)$ and (b) $\delta_{zy}(\omega)$ (dashed lines) for the $40$ wt\% sample.
  The inset in (b) zooms in onto the low-$\omega$ behavior of $G'_{zy}(\omega)$, see also Fig.~\ref{iso_static_m} for the case $\omega=0$.}
  \label{iso_stor_phase_m}
\end{figure}

We now discuss the impact of increasing $M$ on the dynamic moduli.
As expected, in the low-$\omega$ regime ($\omega \lesssim 10^{-1}\unitfreq$) the storage moduli $E'_{zz}(\omega)$ and $G'_{zy}(\omega)$ follow the same behavior as their static counterparts, as shown in Fig.~\ref{iso_stor_phase_m}.
Moreover, the phase lags $\delta_{zz}$ and $\delta_{zy}$ in this regime tend to decrease for increasing $M$.
This implies that a large $M$ does not only increase the storage moduli but also helps to keep stress and strain in phase.
This is consistent because an increase in, e.g., $E'_{zz}$ directly causes a decrease in $\delta_{zz}=\arctan(E''_{zz}/E'_{zz})$.

For larger $\omega\gtrsim 10^{-1}\unitfreq$, the storage Young modulus $E'_{zz}(\omega)$ maintains an overall increasing trend, albeit the amount of increase is partially smaller.
Conversely, the magnetization has a diminishing influence on the phase lag $\delta_{zz}$ in this regime.
For $\omega\gtrsim \unitfreq$ there is no statistically significant variation for $\delta_{zz}$ any longer.
Interestingly, the storage part of the shear modulus $G'_{zy}(\omega)$ decreases when switching on $M$ at intermediate frequencies $\omega$, in contrast to the low-$\omega$ regime.
A crossing of the curves for $G'_{zy}(\omega,M=0)$ and $G'_{zy}(\omega,M=7 \unitmag)$ is observed in Fig.~\ref{iso_stor_phase_m}. % as similarly described in Ref.~\onlinecite{pessot2016dynamic}.
However, for even higher $\omega \gtrsim 10 \unitfreq$ the storage modulus recovers its increasing behavior with increasing magnetization as in the static case.
Approximately, the shear-related  phase lag $\delta_{zy}$ shows a behavior similar to $\delta_{zz}$, although with smaller amplitudes of variation at intermediate frequencies.

\begin{figure}
\centering
  \includegraphics[width=8.6cm]{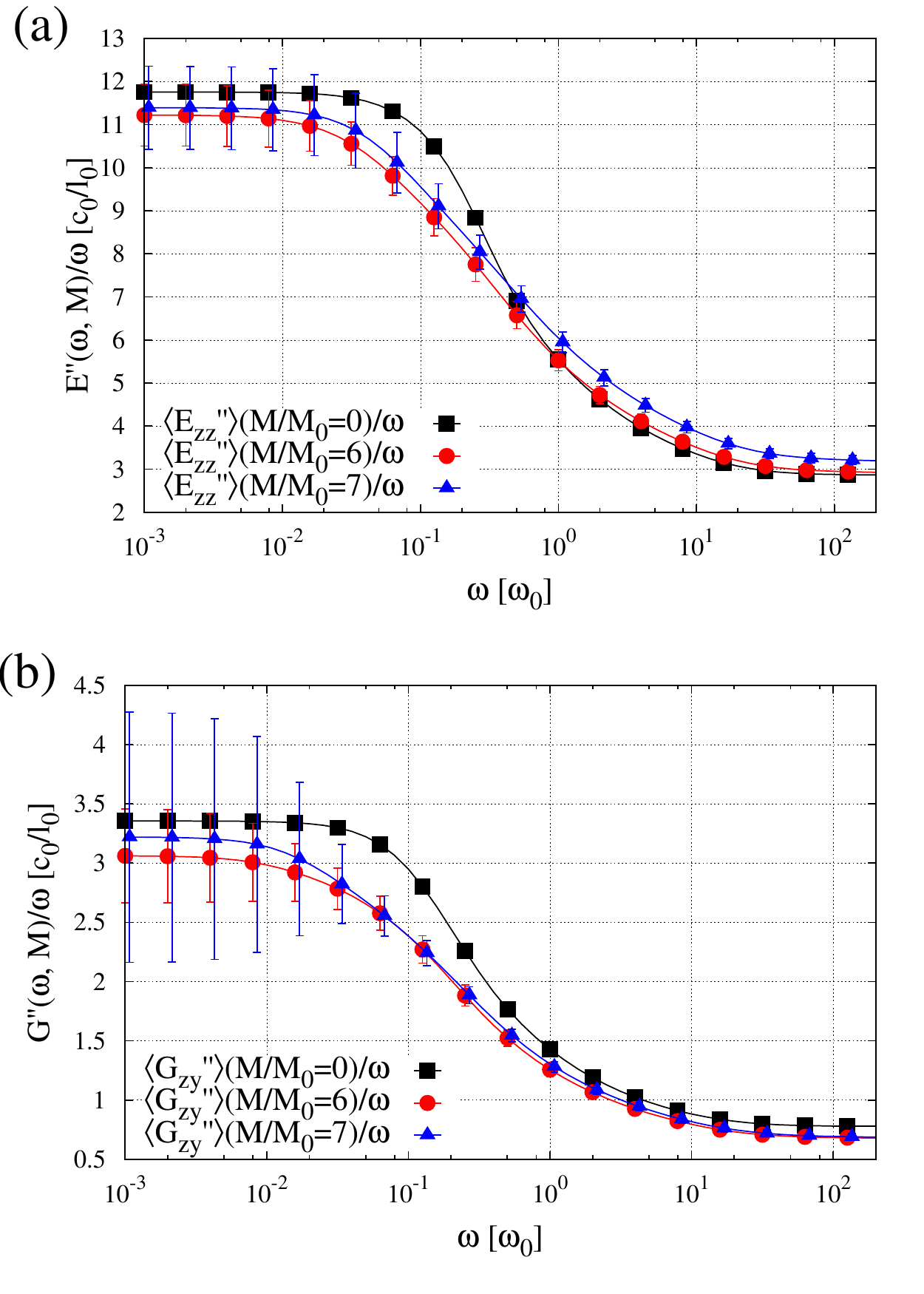}
  \caption{Influence of increasing magnetization $M$ on the loss moduli (a) $E''_{zz}(\omega)$ and (b) $G''_{zy}(\omega)$ for the sample of $40$ wt\%.
  Because of the approximately linear increase of the loss moduli with frequency, we here plot them after division by $\omega$.}
  \label{iso_sloss_m}
\end{figure}

The loss moduli, displayed in Fig.~\ref{iso_sloss_m}, are influenced by increasing magnetization $M$ as well.
Because of the roughly linear behavior in both the low- and high-$\omega$ regime, see Fig.~\ref{lossmods_15_40}, we discuss the variations in terms of changes in $E''_{zz}(\omega)/\omega$ and $G''_{zy}(\omega)/\omega$, which for $\omega\rightarrow 0$ represent the corresponding effective viscosity of the system.

The loss modulus $E''_{zz}(\omega)$ seemingly decreases with increasing $M$ at low to moderate frequencies, thus leading to a reduced effective viscosity.
The reason for the larger standard deviations at low frequencies is that the absolute value of $E''_{zz}(\omega)$ vanishes approximately as $E''_{zz}(\omega) \sim \omega$ for $\omega \rightarrow 0$.
The same applies to the shear loss modulus $G''_{zy}(\omega)$.
While the shear loss modulus $G''_{zy}(\omega)$ always seems to decrease when switching on $M$, see  Fig.~\ref{iso_sloss_m} (b), the Young loss modulus $E''_{zz}(\omega)$ interestingly changes from decrease to increase for higher frequencies.

\begin{figure}
\centering
  \includegraphics[width=8.6cm]{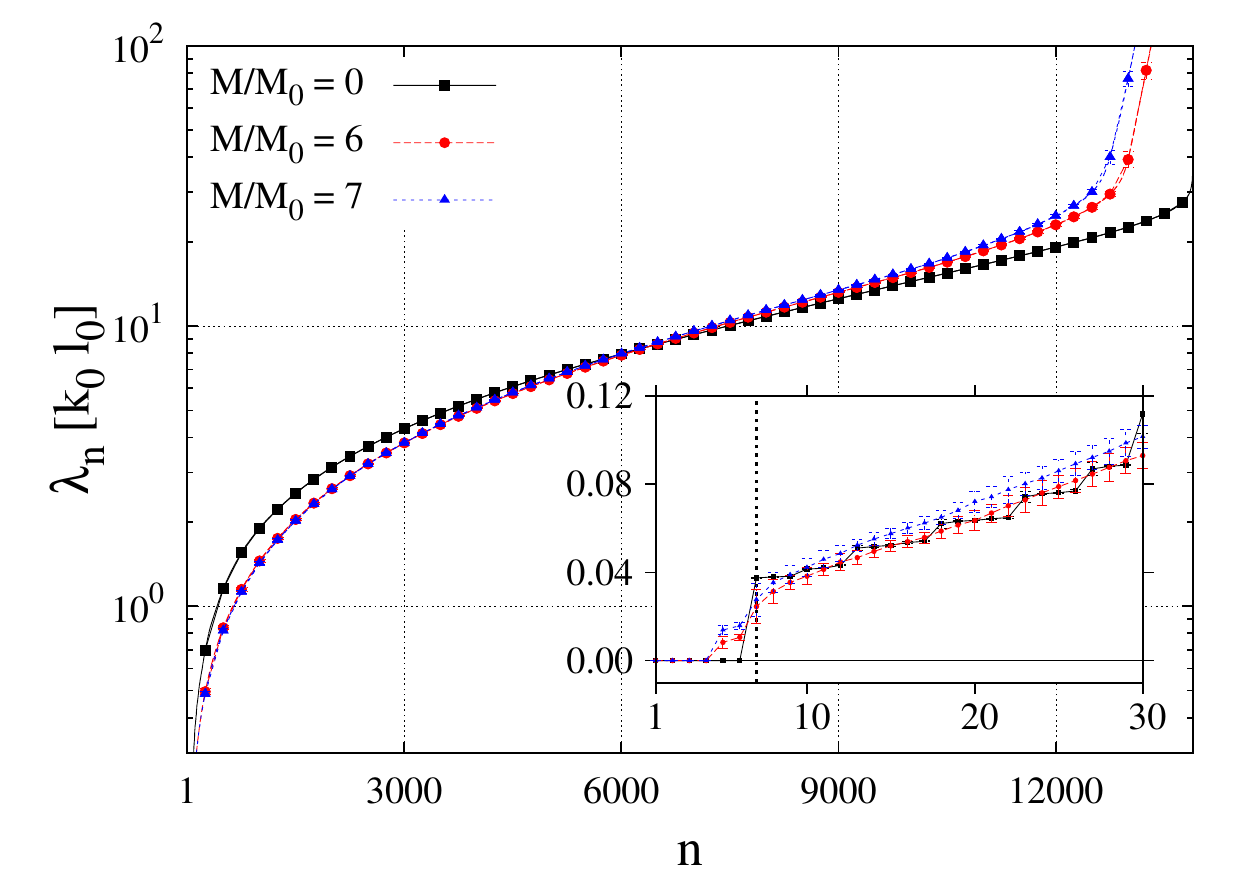}
  \caption{Spectrum of eigenvalues $\lambda_n$ ($n=1\dots 3N$) for the $40$ wt\% sample when increasing the magnetization $M$. %/\unitmag=0,6,7$.
  Data points and bars correspond to averages and standard deviations, respectively, for $\Nsamples$ different numerical realizations of the spring network.
  The inset zooms in onto the lowest $31$ eigenvalues, while the vertical dashed line separates the $6$ lowest $\lambda_n$ representing $3$-dimensional translations and rotations from the other eigenvalues.}
  \label{iso_spectra}
\end{figure}

The chain formation described in section \ref{results_strain_section} and the stiffenings displayed in Figs.~\ref{iso_static_m} and \ref{iso_stor_phase_m} can be related to changes in the distribution of the eigenvalues with increasing $M$, see Fig.~\ref{iso_spectra}.
Over the whole range of $\lambda_n$, the effect of increasing $M$ is to lower the smaller ($\lambda_n \lesssim 10 \unitk \unitl$) and to increase the larger ($\lambda_n \gtrsim 10 \unitk \unitl$) eigenvalues.
This influence is especially pronounced for the larger $\lambda_n$.
They correspond to localized normal modes involving nearby particles.
Such modes typically dominate the response at high frequencies.
The formation of chain-like aggregates strongly magnetically binds the nearby particles to each other, making their relative motion to each other difficult.
This explains the significant increase for larger $\lambda_n$ in Fig.~\ref{iso_spectra} and is directly reflected by the increase of the storage moduli in Fig.~\ref{iso_stor_phase_m} at high frequencies.

%In contrast to that, 
Moreover, the inset of Fig.~\ref{iso_spectra} reveals an increase in many of the very small eigenvalues $\lambda_n$ with increasing magnetization $M$.
At low frequencies, the system has time to significantly adjust to the imposed global deformations.
Therefore, mainly the more globally extended modes corresponding to long-ranged distortions spanning the system become important in the low-$\omega$ regime.
Those are especially the modes corresponding to lowest nonvanishing eigenvalues $\lambda_n$.
Thus, the increase in the elastic storage moduli in Fig.~\ref{iso_stor_phase_m} reflects the increase in these eigenvalues $\lambda_n$ in Fig.~\ref{iso_spectra}.

\section{Conclusions}\label{conclusions_section}
Bringing together experimental analysis of real samples and the subsequent theoretical and numerical investigation and evaluation of the data can complement the two approaches and increase our understanding of complex materials.
Here, refined X-$\mu$CT methods were used to scan macroscopic samples of magnetic elastomers.
\GP{Mesoscopic} information on the positions and volumes of the magnetic particles embedded in an elastic polymeric matrix were obtained this way.
The data were collected for different particle concentrations \cite{gundermann2017statistical, schuemann2017insitu}.
They are then used as input to an \GP{adequately} enhanced version of our recent dipole--spring approach \cite{pessot2016dynamic} of determining the dynamical elastic moduli under varying magnetic interactions.

For this purpose, the elastic polymer matrix in which the particles are embedded is discretized by a randomized network of linear elastic springs.
Each magnetic particle forms a node of the resulting elastic network.
In addition to that, extra nodes not carrying magnetic particles are included to allow for a more homogeneous elastic network.
The particles are assumed of spherical shape, with radii set according to the experimentally measured volumes.
For simplicity, when magnetized, e.g., by an external magnetic field, we assume all particles to show the same magnetization.
Together with the particle volume, it sets the magnetic dipole moment of each particle that leads to magnetic interactions.

As a consequence of this approach, we can theoretically analyze the competition between magnetic and elastic interactions in the experimental particle arrangement for increasing magnetization.
Particularly, this concerns the overall distortion as well as the formation of chain-like aggregates \cite{gundermann2017statistical, schuemann2017insitu, zubarev2016hysteresis, peroukidis2015spontaneous, peroukidis2015tunable}.
Moreover, we can calculate the changes in dynamic elastic storage and loss moduli for increasing magnetization as a function of the frequency of the stress imposed onto the boundaries of the system \cite{pessot2016dynamic}.
Here, we concentrated on a compressive/elongational deformation along the magnetization and a shear deformation containing the magnetization in the shear plane but with boundary displacements perpendicular to it.
Switching on the magnetic interactions, we mostly observe a contraction of the system parallel to the magnetization.
However, the system of higher particle concentration in a certain initial regime also  showed an elongation along this direction.
At high magnetization values, we observe the formation of chain-like aggregates in the system, in which the particles are strongly magnetically bound to each other.
This leads to significant increase in elastic Young and shear storage moduli at low and high frequencies.
Interestingly, a decrease in the elastic shear storage modulus is obtained at intermediate frequencies.
Nonmonotonous behavior as a function of \GP{frequency, when switching on the magnetization}, is also found for the resulting changes in the Young loss modulus.
Additionally, we find an increasing magnetization to reduce the out-of-phase lag between the applied stress and the strain response.

\GP{As mentioned above, a detailed quantitative comparison between the experimental investigation and theoretical results is not possible at the present stage.
However, good qualitative agreement is found with experimental observations accessible so far.
Particularly, the formation of chain-like aggregates under increasing magnetization \cite{gundermann2017statistical, schuemann2017insitu} as well as the increase in the elastic moduli upon chain formation \cite{schuemann2017insitu} have been reported in the experimental investigations.
On the experimental side, a further improvement of the evaluation algorithm will provide increased precision on the particle positions and volumes.
A tracking of the particles for stepwise increase of external magnetic fields and mechanical deformations may further support this effort.
Yet, a simultaneous measurement of structure and dynamics will be very challenging using our set-up.
However, simultaneously measuring structural changes under deformation and quasi-static stress-strain behavior will provide further insight.}
On the theoretical side, an important next task will be to introduce non spherical particle shapes into the formalism.
Moreover, including rotational degrees of freedom \cite{cremer2015tailoring, cremer2016superelastic}, surface-bound springs \cite{cerda2013phase}, induced dipole effects \cite{biller2014modeling, allahyarov2015simulation}, or many-body elastic interactions \cite{puljiz2016forces, puljiz2017forces} will continuously take us towards our goal of combined efforts to study the properties of this class of materials.
In the longer term, also more sophisticated situations such as deformation or actuation of prestretched states \cite{menzel2009response} may be addressed.
In further integrating the experimental and theoretical approaches, we aim at an ongoing process of enhancing the tools to develop and design these appealing materials.

\begin{acknowledgments}
The authors thank the Deutsche Forschungsgemeinschaft for support of this work through the SPP 1681, \GP{grants nos.\ OD 19/2, PAK 907 (OD 18), LO 418/16, and ME 3571/3.}
\end{acknowledgments}

%\textheight=25cm

%\bibliography{litfgels}

\end{document}